\documentclass{aa}
\usepackage{graphicx}

\usepackage{lmodern}
\bibpunct{(}{)}{;}{a}{}{,}
\newcommand{\ha}{H$\alpha$}
\newcommand{\hd}{H$_{2}$}
\newcommand{\nad}{\ion{Na}{I}~D}
\newcommand{\cii}{[\ion{C}{II}]}
\newcommand{\oiii}{[\ion{O}{III}]}
\newcommand{\name}{IRAS~F08572+3915}
\newcommand{\aco}{$\alpha_{\rm CO}$}

\begin{document}

\title{AGN feedback in a galaxy merger: Multi-phase, galaxy-scale outflows including a fast molecular gas blob $\sim6$~kpc away from \name}
\titlerunning{Galaxy-scale Molecular Outflows in a Major Merger}

\author{
	R. Herrera-Camus\inst{\ref{inst1}}
	\and
	A. Janssen\inst{\ref{inst2},\ref{inst3}}	
	\and
	E. Sturm\inst{\ref{inst2}}
	\and
	D. Lutz\inst{\ref{inst2}}
	\and 
	S. Veilleux\inst{\ref{inst4},\ref{inst5},\ref{inst6}}
	\and
	R. Davies\inst{\ref{inst2}}
	\and
	T. Shimizu\inst{\ref{inst2}}
	\and
	E. Gonz\'alez-Alfonso\inst{\ref{inst7}}
	\and
	D. S. N. Rupke\inst{\ref{inst8}}
	\and
	L. Tacconi\inst{\ref{inst2}}
	\and
	R. Genzel\inst{\ref{inst2}}
	\and
	C. Cicone\inst{\ref{inst9}}
	\and
	R. Maiolino\inst{\ref{inst5}}
	\and
	A. Contursi\inst{\ref{inst2}}
	\and
	J. Graci\'a-Carpio\inst{\ref{inst2}}
}
\institute{Astronomy Department, Universidad de Concepci\'on, Barrio Universitario, Concepci\'on, Chile\\
      \email{rhc@astro-udec.cl}\label{inst1}
\and
	Max-Planck-Institute for Extraterrestrial Physics (MPE),
	Gie\ss enbachstra\ss e 1, D-85748 Garching, Germany
	\label{inst2}
\and
	NOVA Optical Infrared Instrumentation Group at ASTRON, P.O. Box 2, 7990 AA, Dwingeloo, The Netherlands \label{inst3}
\and
	Department of Astronomy and Joint Space-Science Institute, Univ. of Maryland, 
	College Park, MD 20742, USA 
	\label{inst4}
\and
	Institute of Astronomy and Kavli Institute for Cosmology Cambridge, University of Cambridge 
	\label{inst5}
\and
	Space Telescope Science Institute, Baltimore, MD 21218
	\label{inst6}
\and
	Departamento de F\'isica y Matem\'aticas, Univ. de Alcal\'a, Campus Universitario, E-28871, Alcal\'a de Henares, Spain 
	\label{inst7}
\and
	Department of Physics, Rhodes College, Memphis, TN 38112, USA 
	\label{inst8}
\and
	Institute of Theoretical Astrophysics, University of Oslo, P.O. Box 1029, Blindern, 0315 Oslo, Norway
	\label{inst9}
}

\abstract{To understand the role that AGN feedback plays in galaxy evolution we need in-depth studies of the multi-phase structure and energetics of galaxy-wide outflows. In this work we present new, deep ($\sim50$~hr) NOEMA CO(1-0) line observations of the molecular gas in the powerful outflow driven by the AGN in the ultra-luminous infrared galaxy \name. We spatially resolve the outflow, finding that its most likely configuration is a wide-angle bicone aligned with the kinematic major axis of the rotation disk. The molecular gas in the wind reaches velocities up to approximately $\pm1200$~km~s$^{-1}$ and transports nearly 20\% of the molecular gas mass in the system. We detect a second outflow component located $\sim6$~kpc north-west from the galaxy moving away at $\sim900$~km~s$^{-1}$, which could be the result of a previous episode of AGN activity. The total mass and energetics of the outflow, which includes contributions from the ionized, neutral, warm and cold molecular gas phases is strongly dominated by the cold molecular gas. In fact, the molecular mass outflow rate is higher than the star formation rate, even if we only consider the gas in the outflow that is fast enough to escape the galaxy, which accounts for about $\sim$40\% of the total mass of the outflow. This results in an outflow depletion time for the molecular gas in the central $\sim$1.5~kpc region of only $\sim3$~Myr, a factor of $\sim2$ shorter than the depletion time by star formation activity.}


\maketitle


\section{Introduction}

Galaxy outflows are practically ubiquitous in the most luminous systems in our nearby universe \citep[e.g.,][]{rhc_heckman00,rhc_rupke05,rhc_veilleux13,rhc_gonzalez-alfonso17,rhc_p-s18}. These outflows encompass multiple gas phases \citep[e.g.,][]{rhc_rupke13,rhc_morganti13,rhc_feruglio15,rhc_fiore17}, are typically fast ($v\gtrsim1000$~km~s$^{-1}$), and can extend over kiloparsec scales. Their properties make them natural candidates for the source of negative feedback required by theoretical models and numerical simulations to quench star formation activity, transforming blue, star-forming galaxies into ``red and dead'' systems \citep[e.g.,][]{rhc_dimatteo05,rhc_hopkins06}. 

While outflow feedback is acknowledged as an important process, the actual physical mechanisms involved are still poorly known. Part of the problem is our limited knowledge of fundamental properties of the outflowing gas such as geometry, multiphase structure, and physical conditions \citep[for a discussion, see][]{rhc_harrison18}. For example, knowledge of the ionized gas density in the outflow is required to convert \ha\ or \oiii\ line luminosities associated with the outflow into ionized gas masses. Electron densities in the outflowing gas around $n_{\rm e}\sim10^{2}$~cm$^{-3}$ are typically assumed, although recent studies suggest that the density could be much higher \citep[$n_{\rm e}\sim10^{3}-10^{5}$~cm$^{-3}$; e.g., ][Shimizu et al. in prep.]{rhc_santoro18,rhc_f-s19,rhc_baron19}, resulting in lower outflow ionized gas masses by up to two or three orders of magnitude. For the molecular phase, studies based on CO lines require an \aco\ factor to convert CO$(1-0)$ luminosities into H$_{2}$ molecular gas masses. \aco\ can vary by a factor of $\sim10$ depending whether the gas is optically-thin or exposed to Galactic excitation conditions. Observational studies tend to favor \aco\ values for molecular outflows between the optically-thin limit \citep[e.g.,][]{rhc_dasyra16} and two to three times the value adopted for ultra-luminous infrared galaxies (ULIRG) \citep[e.g.,][]{rhc_aalto15,rhc_leroy15,rhc_walter17,rhc_cicone18b,rhc_lutz19}. 

Once the outflow gas mass in a given phase is measured, the shape, size and velocity of the outflow are required to calculate the mass outflow rate. This represents another major obstacle as many times observations lack the angular resolution needed to determine the extent and velocity structure of the gas. A common approach to measure mass outflow rates is to assume a spherical or bi-cone geometry with constant velocity. Whether the outflow gas forms a ``thin shell''  \citep[e.g.,][]{rhc_rupke05} or is filled with constant density \citep[e.g.,][]{rhc_maiolino12} leads to mass outflow rate that differ by a factor of three. For a recent discussion on the different ways to measure mass outflow rates depending on the outflow history see \cite{rhc_lutz19}.

Finally, the multiphase nature of galactic outflows implies that measurements of the outflow properties based on a single gas phase can lead to misleading conclusions \citep[for a discussion, see e.g.][]{rhc_cicone18}. Historically, systematic studies of galactic outflows in nearby and high-$z$  galaxies have focused on the ionized gas --observed as broad wing emission in the spectra of the H$\alpha$ and [OIII] lines-- \citep[e.g.,][]{rhc_heckman90,rhc_woo16,rhc_harrison16,rhc_f-s19}, and the atomic phase --based on the Na D or Mg II lines in absorption-- \citep[e.g.,][]{rhc_heckman00,rhc_rupke02,rhc_rupke05,rhc_weiner09,rhc_roberts-borsani19}. The molecular component of outflows, on the other hand, have been much more difficult to study. Great progress was made with the {\it Herschel} Space Observatory using the OH~119~$\mu$m line in absorption to study molecular outflows in Seyfert and luminous infrared galaxies \citep[][]{rhc_fischer10,rhc_sturm11,rhc_veilleux13,rhc_bolatto13,rhc_spoon13,rhc_george14,rhc_stone16,rhc_gonzalez-alfonso17,rhc_zhang18}. More recently, the advent of powerful millimeter-wave interferometers such as the Atacama Large Millimeter/submillimeter Array (ALMA) and the NOrthern Extended Millimeter Array (NOEMA) are rapidly increasing the number of molecular outflows detected based on observations of the CO line \citep[e.g.,][]{rhc_combes13,rhc_sakamoto14,rhc_garcia-burillo14,rhc_leroy15,rhc_feruglio15,rhc_morganti15,rhc_dasyra16,rhc_pereira-santaella16,rhc_veilleux17,rhc_p-s18,rhc_fluetsch19,rhc_lutz19}. At high-$z$, so far only a handful of large-scale, molecular outflows have been studied in QSOs \citep[e.g.,][]{rhc_cicone15,rhc_vayner17,rhc_feruglio17,rhc_carniani17b,rhc_fan18,rhc_brusa18}, sub-millimeter galaxies \citep[e.g.,][]{rhc_spilker18}, and main-sequence, star-forming galaxies \citep[e.g.,][]{rhc_rhc19}.

To understand the existence of kpc-scale molecular outflows produced by active galactic nuclei (AGN) we must first look at the small-scale, mildly relativistic ($\sim0.1-0.3c$) wind driven by AGN radiation pressure \citep[e.g.,][]{rhc_king03,rhc_tombesi15}. This nuclear wind may violently collide with the surrounding interstellar medium (ISM), producing an inner reverse shock that propagates in the rarefied medium and an outer forward-moving shock. In the case of an energy conserving outflow, the shocked gas do not cool, and expands adiabatically. As a consequence, the bulk of the kinetic energy of the wind is transferred to the outflowing gas, which can then expand to reach galaxy-wide scales
\citep[e.g.,][]{rhc_f-g12,rhc_zubovas12,rhc_costa14}.

The fact that most of the mass in large-scale outflows is in the cold, molecular phase \citep[e.g.,][]{rhc_morganti05,rhc_fiore17,rhc_rhc19} is still a matter of study. One alternative is that a large fraction of the hot, outflowing gas is converted into molecular gas due to efficient radiative cooling \citep[e.g.,][]{rhc_zubovas14a,rhc_richings18,rhc_scheneider18}. The other alternative is that cold clouds are driven out of the host galaxy by either thermal-gas ram pressure \citep[e.g.,][]{rhc_tadhunter11, rhc_hopkins12} or radiation pressure from the hotter, outflowing material \citep[e.g.,][]{rhc_murray11,rhc_zhang12}. 

Characterizing the molecular content of outflows is of major importance as molecular gas is the fuel for star formation, thus ejecting a fraction of the molecular gas from the nuclear regions can have a strong impact on their star formation activity. 

To quantify the impact of AGN-feedback on galaxy evolution we require a detailed characterization of the multi-phase structure and energetics of outflows, which in turn requires in-depth studies that minimize the assumptions typically made to estimate key outflow properties such as the mass outflow rate. In that spirit, in this paper we present deep NOEMA CO(1-0) line observations of the molecular gas in the outflow in the ultra-luminous infrared galaxy (ULIRG) IRAS~F08572+3915. These new observations, which achieve an angular resolution a factor of $\sim$2 better compared to previous studies \citep{rhc_cicone14}, improve our constraints on the size, geometry and velocity structure of the outflow, leading to a more reliable measurement of the outflow energetics. Combined with estimates of the outflow properties from other phases (warm molecular, ionized, and atomic), here we present one of the few multi-phase views available of an AGN-driven outflow \citep[for additional examples see][]{rhc_veilleux13,rhc_tombesi15,rhc_feruglio15,rhc_tombesi17,rhc_rupke17}.

\begin{figure*}
\centering
\includegraphics[scale=0.25]{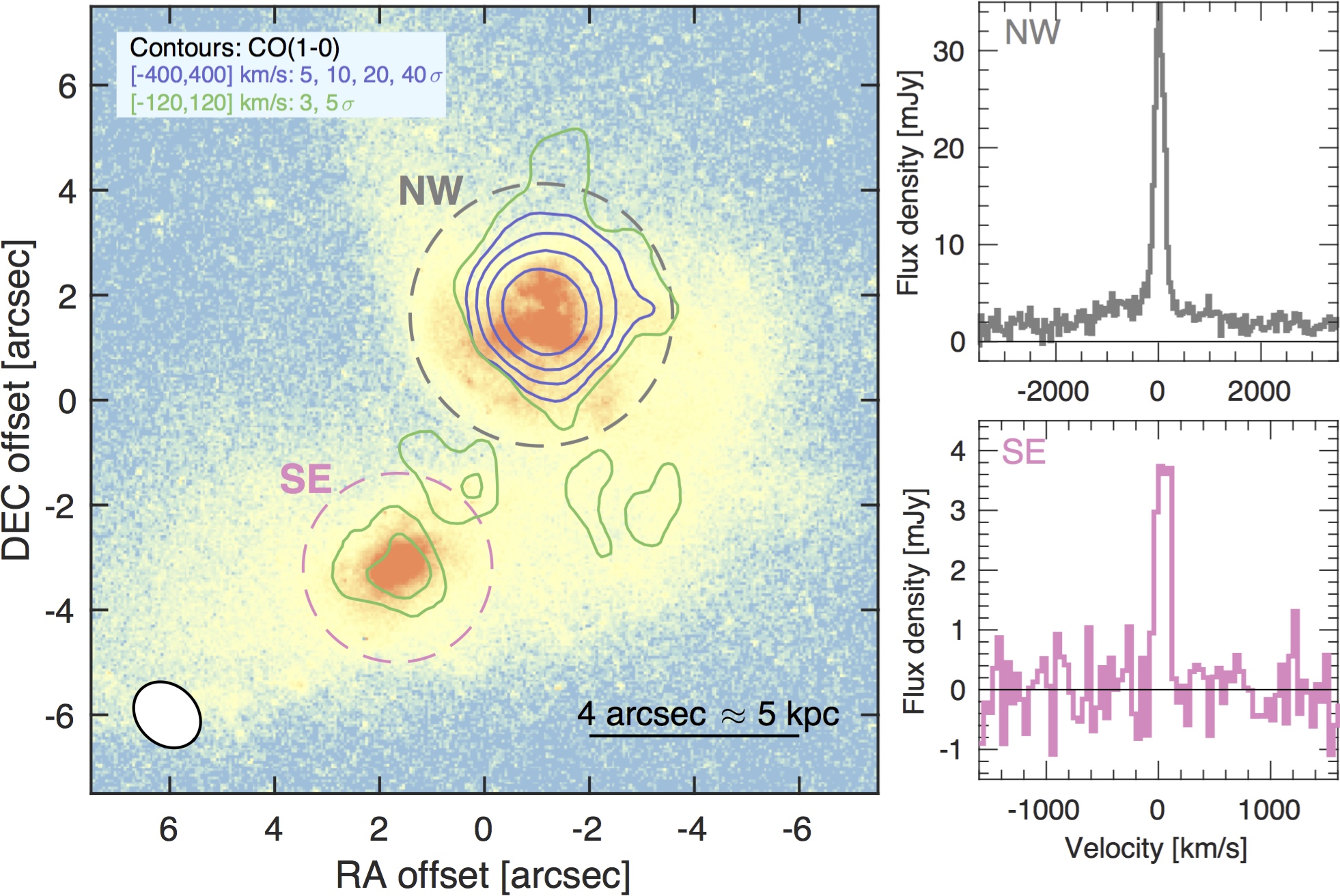}
\caption{{\it (Left)}  HST (F814W) image of the ULIRG IRAS F08572+3915. The system is composed of  a pair of interacting galaxies where the energetics are dominated by a buried AGN in the NW system \citep{rhc_rupke13}. The contours show continuum-subtracted CO(1-0) emission detected at a 3$\sigma$ significance or above in the intensity maps integrated in the [$-$120,+120]~km~s$^{-1}$ (green) and [$-$400,+400]~km~s$^{-1}$ (purple) velocity range. The NOEMA synthesized beam ($\theta=1.4\arcsec\times1.13\arcsec$) is illustrated in the bottom-left corner. The angular resolution achieved is a factor of $\sim2$ better than previous CO(1-0) observations \citep{rhc_cicone14}. {\it (Right)} CO(1-0) spectra of the NW (top) and SE (bottom) galaxies extracted within the circular apertures shown in the left panel. This is the first time the SE component is detected in CO emission.}
\label{fig:hst_co}
\end{figure*}

\subsection{Main properties of IRAS F08572+3915}
\label{sec:properties}

IRAS~F08572+3915 is a low redshift ($z=0.0582$) ULIRG \citep[$L_{{\rm IR},8-1000~\mu {\rm m}}=1.4\times10^{12}~L_{\odot}$;][]{rhc_veilleux13} composed of two interacting spiral galaxies with a separation of about $\sim5$~kpc. Figure~\ref{fig:hst_co} shows an HST (F814W) image of the interacting pair and the circumgalactic material around them. The galaxy located in the north-west quadrant has a stellar mass of $M_{\star}\approx3\times10^{10}$~M$_{\odot}$~yr$^{-1}$ \citep{rhc_rodriguez09} and is $\sim2.5$ magnitudes brighter in K-band than its south-west companion \citep{rhc_scoville00}. Thus, we will refer to this galaxy as the main galaxy in the system.

IRAS F08572+3915 is a key example of a deeply dust-obscured ULIRG with strong mid-infrared silicate absorption \citep[e.g.,][]{rhc_dudley97,rhc_spoon07}. Thus, it is not surprising that strong evidence for AGN activity in the system is only found at infrared wavelengths \citep{rhc_imanishi02,rhc_imanishi06,rhc_armus07}. The system is only marginally detected in soft X-rays \citep{rhc_teng09} and undetected in hard X-rays \citep{rhc_teng15}. In the optical, 
previous classifications of LINER \citep{rhc_veilleux99} or Seyfert~2 \citep{rhc_yuan10} were done based on shallow spectra that shows no clear detection of neither the \oiii~5007$\AA$ nor H$\beta$ lines.

Assuming spherical symmetry, \cite{rhc_veilleux13} estimate that the fraction of the bolometric luminosity of the galaxy ($L_{\rm bol}=1.15L_{\rm IR}$) produced by the AGN is $\alpha_{\rm AGN}=0.74$, which implies an AGN bolometric luminosity of $L_{\rm AGN,bol}=\alpha_{\rm AGN}\times L_{\rm bol}=1.1\times10^{12}~L_{\odot}$, a starburst luminosity of  $L_{\rm SB} = (1-\alpha_{\rm AGN})\times L_{\rm IR}=4.7\times10^{11}~L_{\odot}$, and a SFR of 69~$M_{\odot}~{\rm yr}^{-1}$ \citep[based on the ${\rm SFR}-L_{\rm IR}$ calibration by][]{rhc_murphy11}. This places the main galaxy in \name\ in the ${\rm SFR}-M_{\star}$ plane a factor of $\sim25$ above the main-sequence of galaxies at similar redshift \citep{rhc_whitaker12}. 

Throughout this paper we adopt a cosmology with $H_0=70$~km~s$^{-1}$~Mpc$^{-1}$ and $\Omega_M$ = 0.3, which results in a luminosity distance $D_{\rm L} = 262$~Mpc and a scale of 1.21~kpc/\arcsec\ for a source at $z = 0.0582$.

\begin{table*}
\caption{Details of the NOEMA observations}
\begin{center}
\begin{tabular}{l l l l l}
\hline \hline
Name & Date & Configuration (\# Antennas) & Time on-source & P.I. \\ \hline
 v026 & May - Oct 2011 & C+D (5 or 6) & 20 hr & Sturm \\
 w088 & Feb - March 2013  & A (6) & 10 hr & Sturm \\
 w14ch & March 2015 - Feb. 2016  & A+B (6 or 7) & 20 hr & Janssen \\
\hline
\end{tabular}
\end{center}
\label{tab:obs}
\end{table*}

\section{Observations and Data Reduction}

In total, there have been three IRAM NOEMA (formerly Plateau de Bure Interferometer) observing programs that target the CO(1-0) outflow in IRAS F08572+3915. Table \ref{tab:obs}
lists observing dates, configuration, number of antennas, and on-source time for these programs.  The C+D only data (project v026) was already presented in \cite{rhc_cicone14}. The WideX observations have a band width of 3.6~GHz (corresponding to 9884 km\,s$^{-1}$ at the observed frequency of 108.93~GHz) and a resolution of 1.95~MHz (corresponding to 5.35~km~s$^{-1}$).

The data were calibrated in CLIC with help from the staff in Grenoble. After calibration, separate $uv$ tables were created for the configuration C+D, A+B, and A+B+C+D observations. We then used the software MAPPING2\footnote{CLIC and MAPPING2 part of the GILDAS package \citep{rhc_guilloteau00}: http://www.iram.fr/IRAMFR/GILDAS} for cleaning and imaging in the uv-plane. The synthesized beam size is 2.95\arcsec$\times$2.56\arcsec\ with a Position Angle of 61$^{\circ}$ for the C+D configuration, 1.08\arcsec$\times$0.82\arcsec\ with a P.A. of 34$^{\circ}$ for the A+B configuration, and 1.4\arcsec$\times$ 1.13\arcsec\ with a P.A. of 43$^{\circ}$ for the A+B+C+D configuration. This is a factor of $\sim2$ higher angular resolution than that achieved in the IRAM Plateau de Bure Interferometer (PdBI) observations reported in \cite{rhc_cicone14}. Configuration A+B+C+D observations will be used for further analysis, because of the highest sensitivity. The data were binned in 40~km~s$^{-1}$, which balances a good signal-to-noise ratio and spectral resolution. The continuum is taken to be the average over the velocity range $-$3500~km\,s$^{-1}$ to $-$2000~km\,s$^{-1}$, and 2000~km\,s$^{-1}$ to 4000~km\,s$^{-1}$. It was detected at 1.5~mJy, and has been subtracted from the spectra before any analysis on them was done.

\section{The molecular gas in IRAS F08572+3915}

\begin{figure*}
\centering
\includegraphics[scale=0.27]{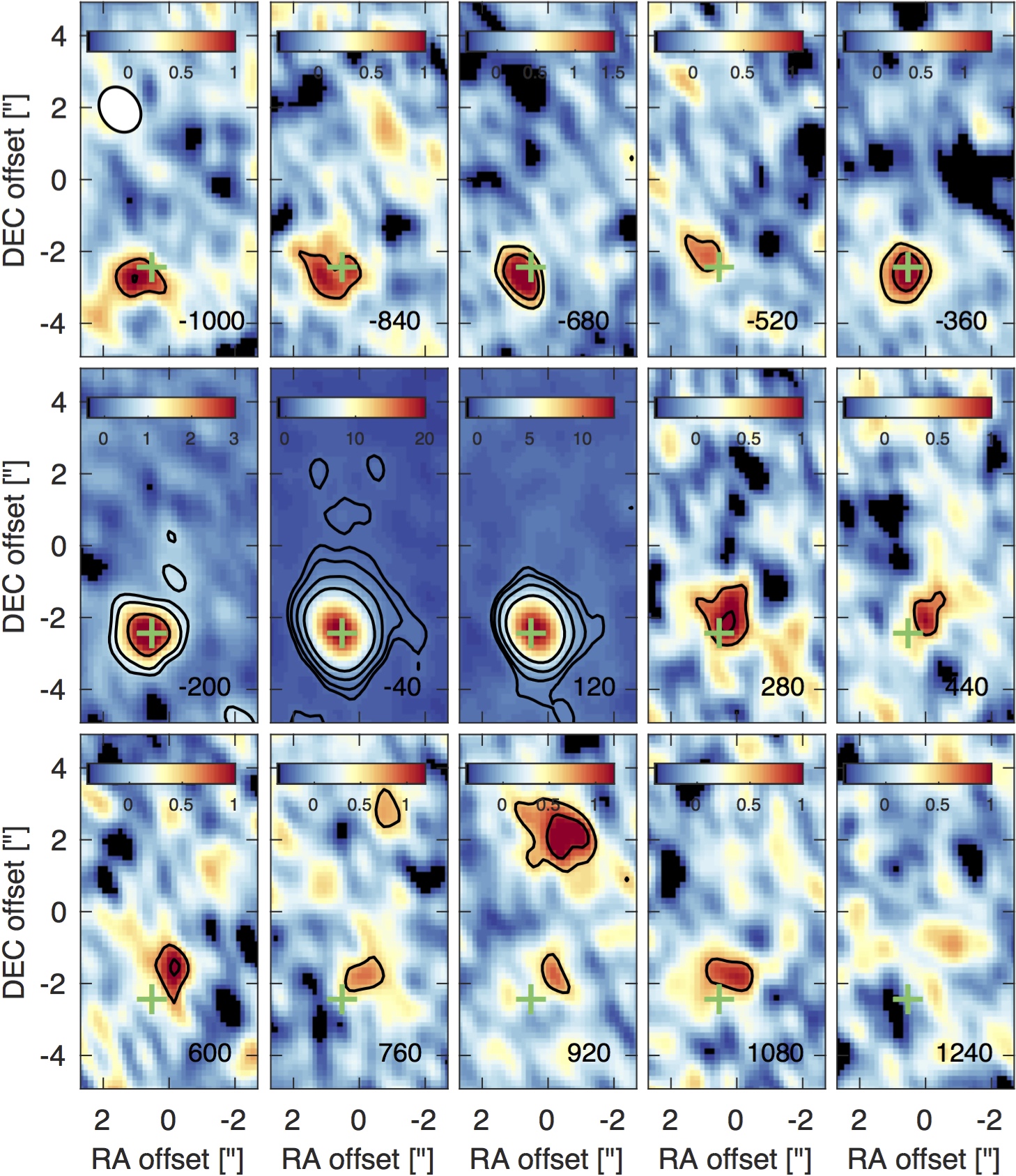}
\caption{Channel maps (40~km~s$^{-1}$ bins) showing the CO(1-0) line emission in IRAS F08572+3915 in velocity steps of 160~km~s$^{-1}$. The velocity of each map is printed in the lower-right corner and the NOEMA beam is shown in the top-left corner of the first panel. The green cross marks the position of the CO(1-0) peak in the NW galaxy, and the contours are placed at the $3, 5, 10$ and $20\sigma$ level. Each map has its own color scale in order to make both bright and faint features visible, so we include the corresponding colorbar in the top of each panel in units of mJy~beam$^{-1}$. We observe that the outflow is aligned with the kinematic major axis of the  disk (roughly going from the south-east to the north-west), and that the  second redshifted outflow component is brightest around 900 km s$^{-1}$.
}
\label{fig:channel_map}
\end{figure*}

\begin{table*}
\caption{CO(1-0) positions, fluxes and masses of the galaxies and outflows}
\begin{center}
\begin{tabular}{cccccccc}
\hline \hline
     &  R.A. &  Dec &  Velocity &   Noise &  Flux &  Molecular Mass &  FWHP \\
     &  9:00:.. &  39:03:.. &  (km s$^{-1}$) &  mJy/beam  &  (Jy km s$^{-1}$) &  ($10^{9}\, M_{\odot}$) & kpc \\
\hline
Blue wing  & 25.40 &  53.9 & $-400$ to $-1200$ & 0.056 & 1.3 & 0.17 & 0.92 \\
Main galaxy & 25.38 & 54.2 & $-400$ to 400 & 0.065& 8.2 & 1.04 & 0.70 \\
Secondary galaxy & 25.6 & 49 & $-400$ to 400 & 0.065 & 0.8 & 0.1 & 1.6 \\
Red wing & 25.32 & 54.8 & 400 to 1200 & 0.055 & 0.9 & 0.1 & 0.98 \\
Red blob & 25.26 &  58.9 & 400 to 1200 & 0.055 & 0.4 & 0.05 & 0.7 \\
\hline                                   
\end{tabular}\\
\end{center}
{Fluxes and gas masses for the individual outflows and galaxies assuming $\alpha_{\rm CO} = 0.8~ M_{\odot} (\text{K}\, \text{km}\, \text{s}^{-1}\, \text{pc}^{2})^{-1}$. All Full Width Half Power (FWHP) values have been derived from the $uv$ data, except for the companion galaxy and the red blob, for which the size is estimated in the image plane.
}
\label{tab:components}
\end{table*}

\subsection{Main galaxy}
\label{sec:nwhost}

The main galaxy in the system (located in the north-west quadrant of Figure~\ref{fig:hst_co}) has previously been observed and detected in CO(1-0) line emission \citep{rhc_solomon97,rhc_evans02,rhc_cicone14}, although with a sensitivity and spatial resolution poorer than the observations presented here. Figure~\ref{fig:hst_co} (left) shows the distribution of the CO emission on top an HST (F814W) image. An elliptical Gaussian fit to the $uv$ table of the A+B+C+D observations gives the peak of emission at R.A. 09:00:25.38 and DEC. $+$39:03:54.2. This position coincides within $0.1\arcsec$ of the radio center at 8.44 GHz found by \citet{rhc_condon91}. The galaxy's emission is not perfectly symmetric around this point, but is more extended towards the West. The best elliptical Gaussian fit has an intrinsic major FWHM of $0.61\pm0.2\arcsec$  and an intrinsic minor FWHM of $0.54 \pm 0.2''$, corresponding to 0.74 by 0.65 kpc, with a P.A. of $-10 \pm 10^{\circ}$. The disk is thus slightly elongated towards the north/north-west, but because the length of the major and minor axes only differ by a little, the P.A. is not used here to constrain the orientation and inclination of the disk.

The top-right panel of Figure~\ref{fig:hst_co} shows the CO spectrum of the main galaxy. We measure a redshift of $z_{\rm CO}=0.0582$, which we use to set the systemic velocity. This redshift is similar to that found by \citet{rhc_evans02} and the same as the one measured by \cite{rhc_gonzalez-alfonso17} based on the [CII] line. The continuum level, depth of observations, and resolution do not indicate any absorption related to the wind seen by \cite{rhc_geballe06} and \cite{rhc_shirahata13} in other wavelengths.

\begin{figure}
\centering
\includegraphics[scale=0.23]{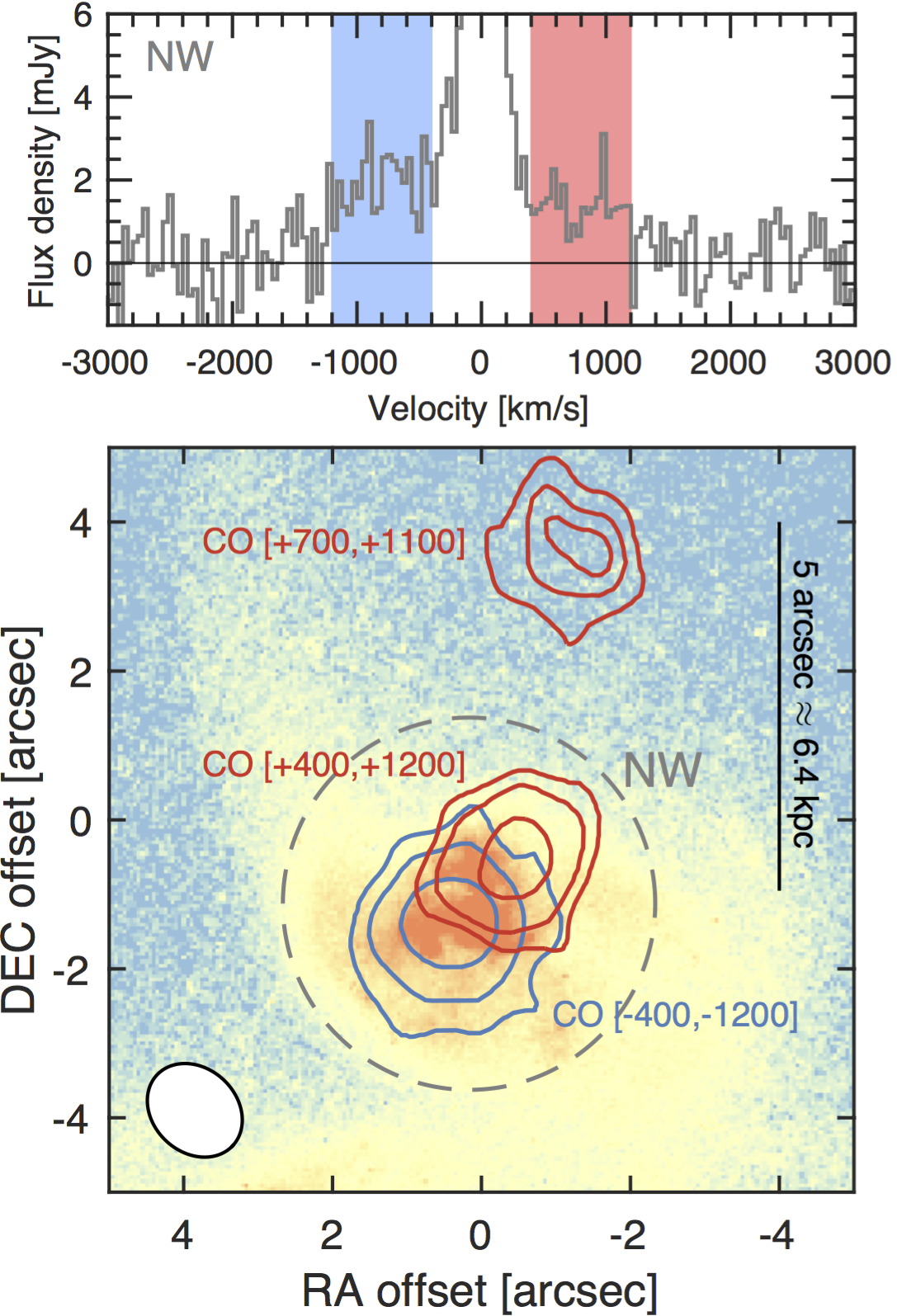}
\caption{{\it (Top)} Continuum-subtracted CO(1-0) spectrum extracted within a circular aperture of 5\arcsec diameter centered on the NW galaxy. Strong broad wings of CO(1-0) line emission indicative of a molecular outflow are detected with velocities up to $\pm$1200~km~s$^{-1}$. {\it (Bottom)} Integrated CO(1-0) line emission integrated in the blue ($[-400,-1200]$~km~s$^{-1}$) and red ($[+400,+1200]$~km~s$^{-1}$) wings and the red gas blob ($[+700,+1100]$~km~s$^{-1}$) overplotted on a HST (F814W) image. The NOEMA synthesized beam is shown in the bottom-left corner. Contours correspond to 3, 5, and 10$\sigma$ for the blue and red wings, and 3, 5, and 7$\sigma$ for the red gas blob. An extranuclear component of the red wing is detected approximately $\sim$6~kpc north (projected distance) of the NW galaxy moving at $\sim+900$~km~s$^{-1}$, but decelerating (see Fig.~\ref{fig:co_moments}).
}
\label{fig:co_wings}
\end{figure}
 
\begin{figure*}
\centering
\includegraphics[scale=0.25]{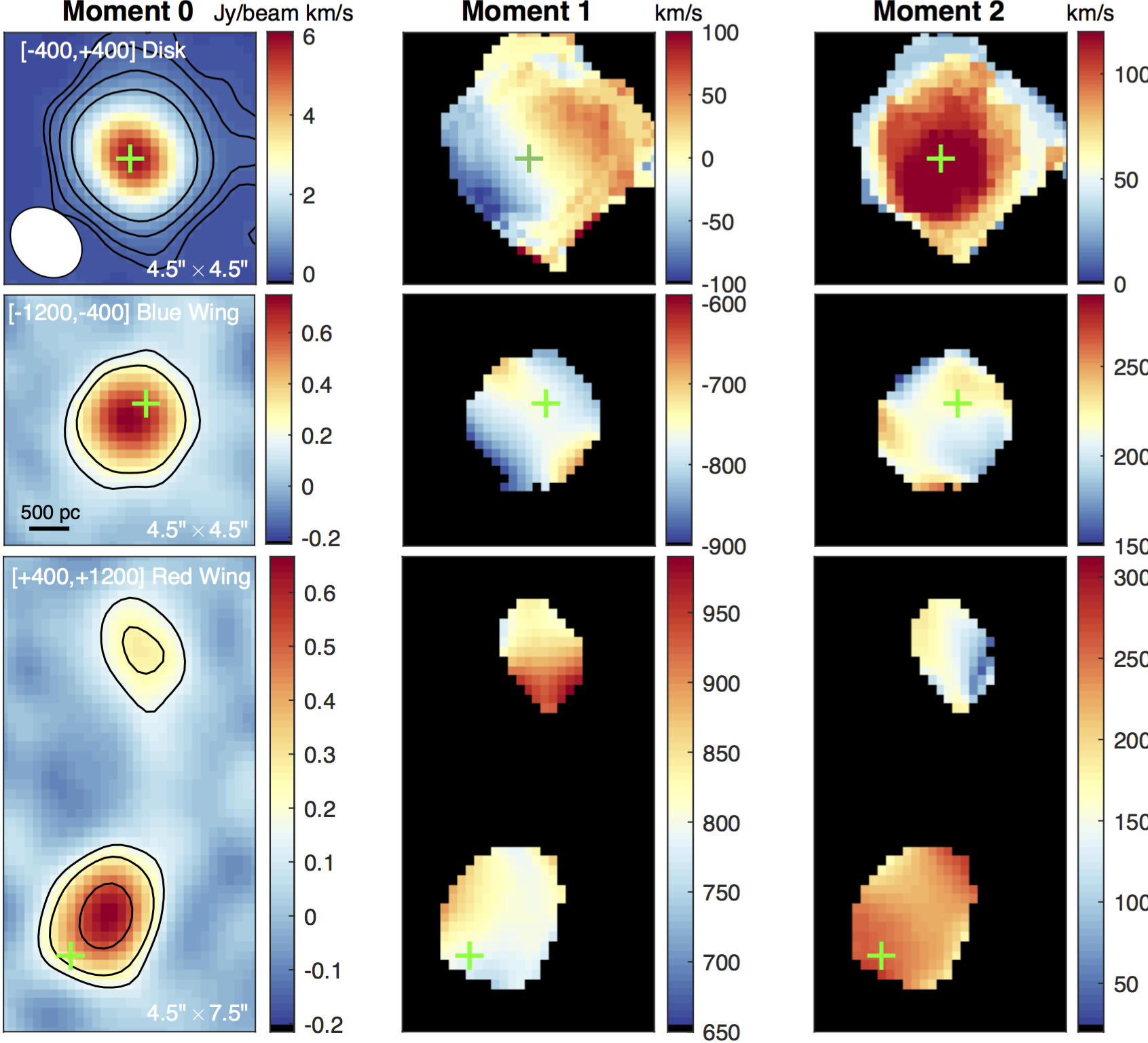}
\caption{Integrated intensity (left column), velocity (center column) and dispersion (right column) maps for the NW galaxy (top row), the blue wing (center row), and the two red wing components (bottom row). Only  regions with $> 3\sigma$ detections are shown. The green cross marks the position of the CO(1-0) peak in the NW galaxy. Contours correspond to 3, 5, 10 and 20$\sigma$, where $\sigma$ for each map is listed in Table~\ref{tab:components}. The NOEMA beam is shown in the bottom-left corner of the first panel.}
\label{fig:co_moments}
\end{figure*}

From visual inspection, we decided to measure the flux in the galaxy by integrating in the $[-400,+400]$~km~s$^{-1}$ velocity range. This results in $F_{\rm CO(1-0)}=8.2\pm0.4$~Jy~km s$^{-1},$\footnote{Here we assume a 5\% flux calibration uncertainty. According to the ``IRAM Plateau de Bure Interferometer Data Reduction Cookbook'' \citep{rhc_castro10}, the flux calibration accuracy at 3~mm is $\lesssim10\%$.} which is consistent with the single-dish (IRAM 30~m) measurement of $F_{\rm CO(1-0)}=9.0\pm1.8$~Jy~km s$^{-1}$ by \cite{rhc_solomon97}. Our CO flux measurement corresponds to a molecular gas mass of $M_{\rm mol}=1.04\pm0.10\times10^9$~M$_{\odot}$ assuming a ULIRG-like conversion factor of  $\alpha_{\rm CO,ULIRG}=0.8~{\rm M}_{\odot}~({\rm K~km~s^{-1}~pc}^{2})^{-1}$ \citep{rhc_downes98}. However, more recent studies by \cite{rhc_genzel15} and \cite{rhc_tacconi18} that compare CO and dust-based estimates of the molecular gas mass in nearby and high-$z$ galaxies suggest that the $\alpha_{\rm CO}$ factor applied to (U)LIRGs should be the standard Milky Way conversion factor of $\alpha_{\rm CO,MW}=4.4~{\rm M}_{\odot}~({\rm K~km~s^{-1}~pc}^{2})^{-1}$ \citep{rhc_bolatto13}. Assuming the latter yields a molecular gas mass of $M_{\rm mol}=5.72\times10^9$~M$_{\odot}$. Relative to the stellar mass of the main galaxy of $M_{\star}\approx3\times10^{10}$~M$_{\odot}$ \citep{rhc_rodriguez09}, the molecular gas-to-stellar mass ratio is $\mu\equiv M_{\rm mol}/M_{\star} \approx 3\%$ or $\approx16\%$ if we assume  $\alpha_{\rm CO,ULIRG}$ or  $\alpha_{\rm CO,MW}$, respectively.

The position, velocity range, flux, molecular gas mass, and size of the main galaxy are listed in Table \ref{tab:components}.

\subsection{Companion galaxy}
\label{sec:swcomp}

The left panel of Figure~\ref{fig:hst_co} shows extended CO emission at the location of the companion galaxy in the south-east quadrant. This is the first time this galaxy is detected in CO line emission. The peak is at R.A. 9:00:25.6 and Dec. $+$39:03:49, and coincides with the position of the galaxy in SDSS $i-$band images and the HST (F814W). The line profile has a regular shape, is narrow ($\sigma=59$~km~s$^{-1}$), and peaks at $v= 30$~ km s$^{-1}$ (bottom-left panel in Figure~\ref{fig:hst_co}). The line has a flux of $F_{\rm CO(1-0)}=0.8$~Jy~km~s$^{-1}$, which corresponds to a molecular gas mass of $M_{\rm mol}\approx 10^{8}\,M_{\odot}$ assuming a conversion factor $\alpha_{\rm CO,ULIRG}$ (Table \ref{tab:components}) and $M_{\rm mol}\approx 5\times10^{8}\,M_{\odot}$ assuming $\alpha_{\rm CO,MW}$. There is a substantial uncertainty in the flux (estimated to be $\sim 20\%$), because the emission is extended and contaminated by residual side lobes from the main source. A Gaussian fit to the  emission in the image plane results in a FWHM of $1.8\arcsec$. With an average beam size of $1.25\arcsec$, the estimated (deconvolved) FWHP of the SE galaxy is $1.3\arcsec$ or 1.6 kpc. The size has been fit in the image plane, because a fit in $uv$ plane was not successful.

\section{The outflow in IRAS F08572+3915}
\label{sec:biconical}

As the top panels in Figure~\ref{fig:hst_co} and Figure~\ref{fig:co_wings} show, the spectrum of the main galaxy shows clear evidence for high-velocity gas material that extends up to velocities of $\pm1200$~km~s$^{-1}$. These broad wings of emission are strongly suggestive of the presence of a fast molecular outflow. 

\subsection{Spatial distribution and kinematics of the outflow}

The spatial and velocity distribution of the molecular gas can be further explored by looking at the channel maps in Figure~\ref{fig:channel_map}. The first four panels ($v_{\rm CO}\lesssim-400$~km~s$^{-1}$) reveal the blueshifted wing of the outflow. This component is located south-east of the main galaxy and is centered at roughly the same position in all four bins; no evident velocity gradient is visible  within our angular resolution. The next five channels encompass the velocity range $-400\lesssim v_{\rm CO}\lesssim+400$~km~s$^{-1}$ where the bulk of the CO emission arises from the body of the main galaxy.  Finally, the bottom panels reveal the redshifted component of the outflow ($v_{\rm CO}\gtrsim+400$~km~s$^{-1}$) located north-west of the galaxy center. Similar to the blue wing, this component remains centered at a similar position, except for the channel at $v_{\rm CO}=920$~km~s$^{-1}$. Around this velocity a second outflow component is present located at about $\sim6$~kpc from the main galaxy in the northern direction. 

A complementary view of the outflow structure is provided in Figure~\ref{fig:co_wings}, which shows the spatial distribution of the integrated CO emission in the blue ($[-1200,-400]$~km~s$^{-1}$) and red ($[+400,+1200]$~km~s$^{-1}$) wings of the spectrum overplotted on a HST (F814W) image. The contours are placed at $3, 5~{\rm and}~10\sigma$ levels, with $1\sigma$ corresponding to the noise level given in Table \ref{tab:components}. As already revealed in the channel maps, two main outflows components can be identified: (1) a main component centered at the position of the NW galaxy, and (2) a fast ``blob'' of gas located $\approx 5''$ ($\approx6.4$~kpc) north from the center. This second, fainter, outflow gas has been detected before by \citet{rhc_cicone14}, and now thanks to the higher angular resolution of our observations, we confirm this is a different component.

Finally, Figure~\ref{fig:co_moments} shows the integrated intensity, velocity, and velocity dispersion maps of the main galaxy, and the blue and red components of the outflow. The velocity map of the main component galaxy shows a convincing but slightly disturbed rotating disk with a position angle of approximately $-45^{\circ}$ degrees. The blue and red main outflow components have velocity dispersions in the $\sigma_{\rm CO}\approx150-300$~km~s$^{-1}$ range, and velocity fields that show no systematic variations as a function of position. In contrast, the velocity of the red gas ``blob" decreases from $\sim$1000~km~s$^{-1}$ to about $\sim$850~km~s$^{-1}$ as a function of increasing distance from the host.

We discuss in more detail the nature of this second outflow component in Sections~\ref{sec:second_red_outflow} and \ref{sec:origin_blob}. 

\subsection{Geometry of the outflow}
\label{sec:geometry}

The observed spatially-resolved properties of the outflow --including the absence of a velocity gradient (within our resolution), the large range of velocities covered, and the spatial offset between the main outflow components and the galaxy center-- provide useful constraints on its structure.

We consider two possible ideal cases: a bicone (a shell with an opening angle), and two individual blobs. In both cases, the outflow has a maximum velocity $v_{\text{max}}$, and is only slightly resolved spatially.

\begin{itemize}
 \item In the case of a bicone, if the angle is large enough so the geometry approaches a shell, or if the bicone is pointed directly toward the observer, then we would expect that the channel maps at all velocities to be centered at the same point. In any other case we would expect that the blueshifted and redshifted emission should be offset from the center of the outflow, in opposite directions. Acceleration and deceleration may be observed, depending on the opening angle.
 \item In the case of individual clouds, the observed velocity range is caused by turbulence within the clouds, rather than projection effects. Unless the cloud moves directly toward us, it should be clearly offset from the driving source. Velocity gradients could be observed when the cloud accelerates or decelerates.
\end{itemize}

The main outflow component matches best with the description of the biconical outflow with a large opening angle, directed not exactly toward us but making an angle with the line-of-sight. Because of the large opening angle, we expect $v_{\text{max}}$ to be close to the maximum observed velocity in the outflow, which is $\sim$1200~km~s$^{-1}$. The second redshifted outflow, on the other hand, matches the description of the individual cloud. The assumed geometry of the outflow has implications for the calculation of the mass loss rate and energetics as we discuss in the next section. 

\begin{table*}
\caption{Positions, fluxes and masses of the galaxies and outflows}
\begin{center}
\begin{tabular}{lccccccc}
\hline \hline
 Tracer &  Mass &  $v$ &  Radius &  $dM/dt$ &   $c \times v \times dM/dt$ &  $1/2\times v^{2}dM/dt$ & Reference \\
        &  $M_{\odot}$ &  km/s   &  kpc     &  $M_{\odot}$/yr &  $10^{12} \times L_{\odot}$ &  $10^{43}$ erg/s &  \\
   (1)    &  (2)           &  (3)      &  (4)       &  (5)                &  (6)     &  (7)               &  (8) \\ \hline
  H$\alpha$ & $8.5 \times 10^{6}$ & 1524 &  $ \leq 2$ &7.6 & 0.69 & 1.3 & a \\
  Na I D & $7.6 \times 10^{7}$ & 403 &  $\leq 2$ & 24.5 & 0.89 & 0.63 & a \\
  OH     & $1.2 \times 10^{8}$ & 500-950 &  0.11 &650 & 22 & 10  & b \\
  H$_{2}$ & $5.2 \times 10^{4}$  & 1000 &  0.4 & 0.13 & 0.006 & 0.004 &  c \\
  $\lbrack$CII$\rbrack$ & $1.4 \times 10^{8} $ & 800 & ... & ... & ... & ... & d \\
  CO(1-0)$_{\text{CD}}$ & $4.1 \times 10^{8}$ & 800 &  0.82 & 1210 & 47.8 & 24.5 & e \\
  CO(1-0)$_{\text{ABCD}}$ & $2.7 \times 10^{8}$ & 1200 &  0.95 & 350 & 21 & 16 & this work \\
\hline
\hline
\end{tabular}\\
\end{center}
{Outflow properties of different ISM phases. As a comparison, the SFR is 69~$M_{\odot}$/year \citep{rhc_gonzalez-alfonso17}. Columns: (1) Tracer, (2) Total gas mass in the outflow, (3) average or typical outflow velocity, (4) radius in kpc, (5) Outflow mass loss rate,  (6) outflow momentum rate, (7) kinetic power in the outflow. References: (a)  \citet{rhc_rupke13}, (b) \cite{rhc_gonzalez-alfonso17}, (c) \citet{rhc_rupke13b}, (d) \citet{rhc_janssen16}, (e) \citet{rhc_cicone14}. The differences between the newest CO observations (CO(1-0)$_{\text{ABCD}}$) and those presented earlier (CO(1-0)$_{\text{CD}}$) are mostly caused by new insights in the outflow geometry. See \S \ref{sec:outflow_size} and \S\ref{sec:other_phases} for details.}
\label{tab:phases}
\end{table*}

\subsubsection{Size, mass outflow rate, and energetics}
\label{sec:outflow_size}

The sizes of the two components in the biconical outflow are retrieved from Gaussian fits to the uv data. The Full Width at Half Power (FWHP) of the blueshifted part is $0.77''$ or 0.92 kpc, in comparison to the 1.36 kpc found in \citet{rhc_cicone14}: adding visibilities at larger $uv$ radii resulted in a better fit with a slightly more compact source. For the redshifted component, the difference between our FWHP and that found in \citet{rhc_cicone14} is larger because the second redshifted outflow is now resolved, and not included in the Gaussian fit. This results in a FWHP of the redshifted part of $0.82''$ or 0.98 kpc (compared to 1.91 kpc found previously). The FWHP is listed in Table~\ref{tab:components}.

Since the size of the red outflow component ($\sim0.9$~kpc) is smaller than its distance to the galaxy center ($\sim1.1$~kpc), the outflow seems detached from the galaxy, as if it is not replenished with new gas (see also the channel maps in Figure \ref{fig:channel_map}). This observation suggests {\it that the outflow is a bursty rather than a continuous process. }  This has consequences for the calculation of the outflow mass. \citet{rhc_maiolino12} derive the mass outflow rate for a spherical outflow with uniform density that is continuously replenished with new gas to be $\dot{M}_{\text{out}} = 3M_{\rm out}v_{\text{max}}/R$. This changes to an instantaneous mass outflow rate of $\dot{M}_{\text{out}} = M_{\rm out}v_{\text{max}}/\Delta R$ for a bursty outflow. In the latter case, the thickness of the outflowing shell, $\Delta R$, is of interest. This value is hard to derive observationally in a biconical outflow, but the FWHP as given in Table~\ref{tab:components} is the best approximation. Assuming then an $\alpha_{\rm CO,ULIRG}$ conversion factor for the gas in the outflow, $v_{\rm out}=1200$~km~s$^{-1}$ (see \S\ref{sec:geometry}), and $\Delta R=0.95$~kpc (the average FWHP between the red and blue wings; see Table~\ref{tab:components}), the molecular mass outflow rate in IRAS~F08572+3915 is:

\begin{equation}
\begin{split}
\dot{M}_{\rm out,mol}\approx350~M_{\odot}~{\rm yr}^{-1}\times\bigg(\frac{\alpha_{\rm CO,out}}{\alpha_{\rm CO,ULIRG}}\bigg)\\
\times\bigg(\frac{v_{\rm out}}{1200~{\rm km}~{\rm s}^{-1}}\bigg)\times\bigg(\frac{~{\rm 0.95~kpc}}{\Delta R}\bigg).
\end{split}
\end{equation}

\noindent Since $\Delta R$ is probably overestimated, this mass outflow rate represents a conservative estimate. This value is also $\sim3$ times smaller than that derived previously by \cite{rhc_cicone14}. This is mainly caused by two changes: (1) the second redshifted outflow is excluded from the analysis, resulting in a smaller outflow mass by $\sim15\%$ (see Figure~\ref{fig:co_wings}), and (2) the outflow rate is calculated as $\dot{M}_{\text{out}} = Mv_{\text{max}}/\Delta R$ rather than as $\dot{M}_{\text{out}} = 3\times Mv_{\text{max}}/R$, based on new insights that the outflow is bursty rather than continuous.

In addition to the mass outflow rate, we can also calculate the momentum ($\dot{P}_{\rm out} = \dot{M}_{\rm out} \times v_{\rm out}$) and energy flux ($\dot{E}_{\rm out} = \frac{1}{2} \dot{M}_{\rm out} \times v_{\rm out}^2$) of the molecular outflow in IRAS~F08572+3915. Based on the values listed in Table~\ref{tab:components}, the outflow momentum flux in IRAS~F08572+3915 is

\begin{equation}
\begin{split}
\dot{P}_{\rm out,mol}\approx 2.7\times10^{36}~{\rm dynes}\times\bigg(\frac{\alpha_{\rm CO,out}}{\alpha_{\rm CO,ULIRG}}\bigg)\\
\times\bigg(\frac{v_{\rm out}}{1200~{\rm km}~{\rm s}^{-1}}\bigg)^2\times\bigg(\frac{0.95~{\rm kpc}}{R_{\rm out}}\bigg)
\end{split}
\end{equation}

\noindent and the outflow kinetic energy flux is

\begin{equation} \label{eq:E}
\begin{split}
\dot{E}_{\rm out,mol}\approx1.6\times10^{44}~{\rm erg~s}^{-1}\times\bigg(\frac{\alpha_{\rm CO,out}}{\alpha_{\rm CO,ULIRG}}\bigg)\\
\times\bigg(\frac{v_{\rm out}}{1200~{\rm km}~{\rm s}^{-1}}\bigg)^3\times\bigg(\frac{0.95~{\rm kpc}}{R_{\rm out}}\bigg).
\end{split}
\end{equation}

We analyze these quantities in the context of the momentum boost and the power generated by the starburst and the AGN activity present in IRAS~F08572+3915 in Section~\ref{sec:power_source}.

\subsection{The second redshifted outflow $\sim$6~kpc north of the main galaxy}
\label{sec:second_red_outflow}

The second and independent part of the redshifted outflow is located at R.A. 9:00:25.26, Dec. $+$39:03:58.9, at $4.9''$ or 5.9~kpc north from the main galaxy. The spectrum extracted within a $R=1.5\arcsec$ circular aperture centered at this position is shown in Figure~\ref{fig:gasblob_spec} (this outflow component is also detected in the A+B only array configuration data, see Appendix~\ref{blob_spec_AB}).
It has a flux of $F_{\rm CO(1-0)}=0.4$~Jy~km~s$^{-1}$, corresponding to a molecular gas mass of $M_{\rm mol}\approx 5 \times 10^{7}$~M$_{\odot}$ (assuming a conversion factor $\alpha_{\rm CO,ULIRG}$). No optical counterparts for this outflow have been found. The size of the outflow cannot be retrieved from a Gaussian fit in the $uv$-plane, because it is faint. We therefore estimate the size from a Gaussian fit in the image plane, deconvolved with the beam size. The resulting projected size (FWHM) is 0.7~kpc.

Although this outflow component is barely resolved, it seems to have a velocity gradient, ranging from 1000~km~s$^{-1}$ on the side facing the galaxy to about 600~km~s$^{-1}$ on the opposite side (see Figure~\ref{fig:channel_map} and \ref{fig:co_moments}). The gradient is seen in all channel maps, regardless of their bin size. Moreover, the shift is observed in spectra taken at different distances from the host, which reinforces the idea that the velocity gradient is real. Since there is no clear obstacle in the way of the outflow, the gas might simply slow down as a result of gravitational pull and the lack of an ongoing driving mechanism, or because there is more energy being deposited closer to the nucleus in the NW galaxy. We discuss possible scenarios for the origin of the gas blob in Section~\ref{sec:origin_blob}.

\begin{figure*}
\centering
\includegraphics[scale=0.065]{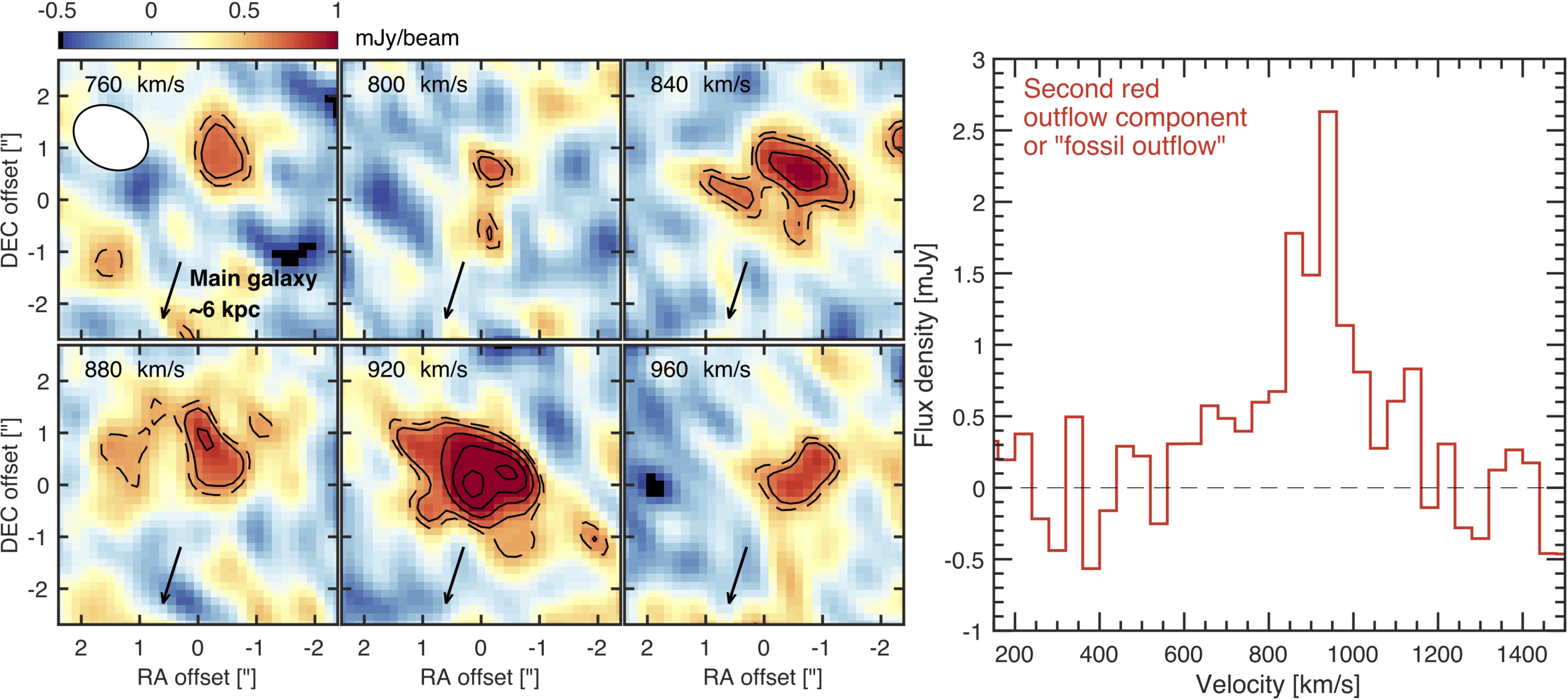}
\caption{(Left) Channel maps showing the CO(1-0) line emission of the ``gas blob'' (R.A. 9:00:25.26, Dec. $+$39:03:58.9) located $\sim$6~kpc north of the main galaxy. The NOEMA beam is shown in the top-left corner of the first panel. The contours show the 2.5$\sigma$ (dashed line) and 3, 4, 5 and 6$\sigma$ (solid lines) levels of emission. (Right) CO(1-0) spectrum extracted from a circular region with a radius of 1.5\arcsec\ centered around the ``gas blob''.
}
\label{fig:gasblob_spec}
\end{figure*}

\begin{figure}
\centering
\includegraphics[scale=0.14]{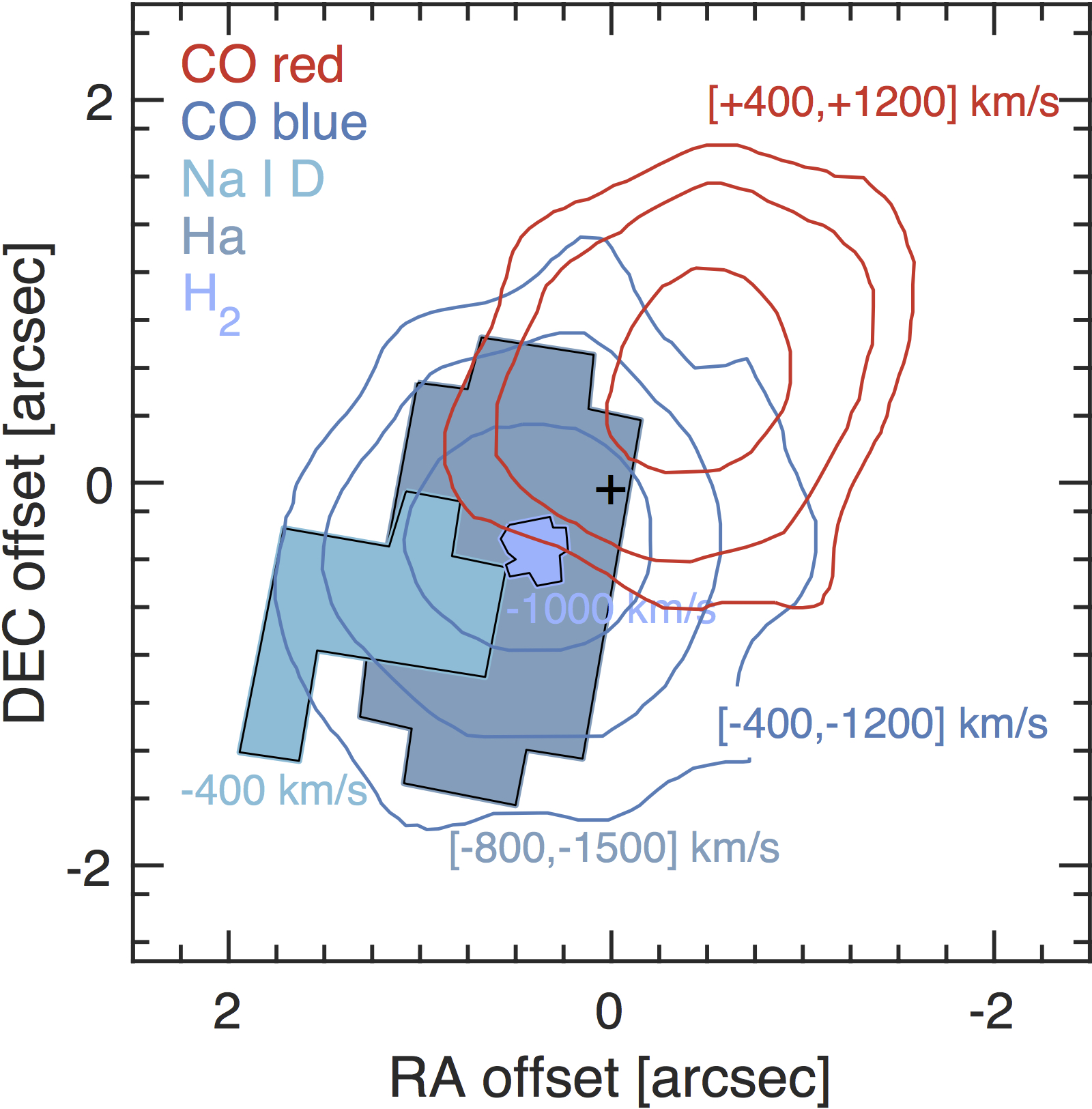}
\caption{
The multiphase structure of the outflow in IRAS F08572+3915, including the cold molecular wind as traced by the CO(1-0) line (this work; blue and red contours), and the warm molecular (\hd), atomic (\nad), and ionized winds (\ha) as reported by \cite{rhc_rupke13}. The black cross marks the position of the CO(1-0) peak in the NW galaxy. The observed range of velocities in the winds is listed next to each outflow component. Absorption by the galaxy's disk is probably the reason that the redshifted outflow is only detected in CO.}
\label{fig:co_multi}
\end{figure}

\begin{figure*}
\centering
\includegraphics[scale=0.23]{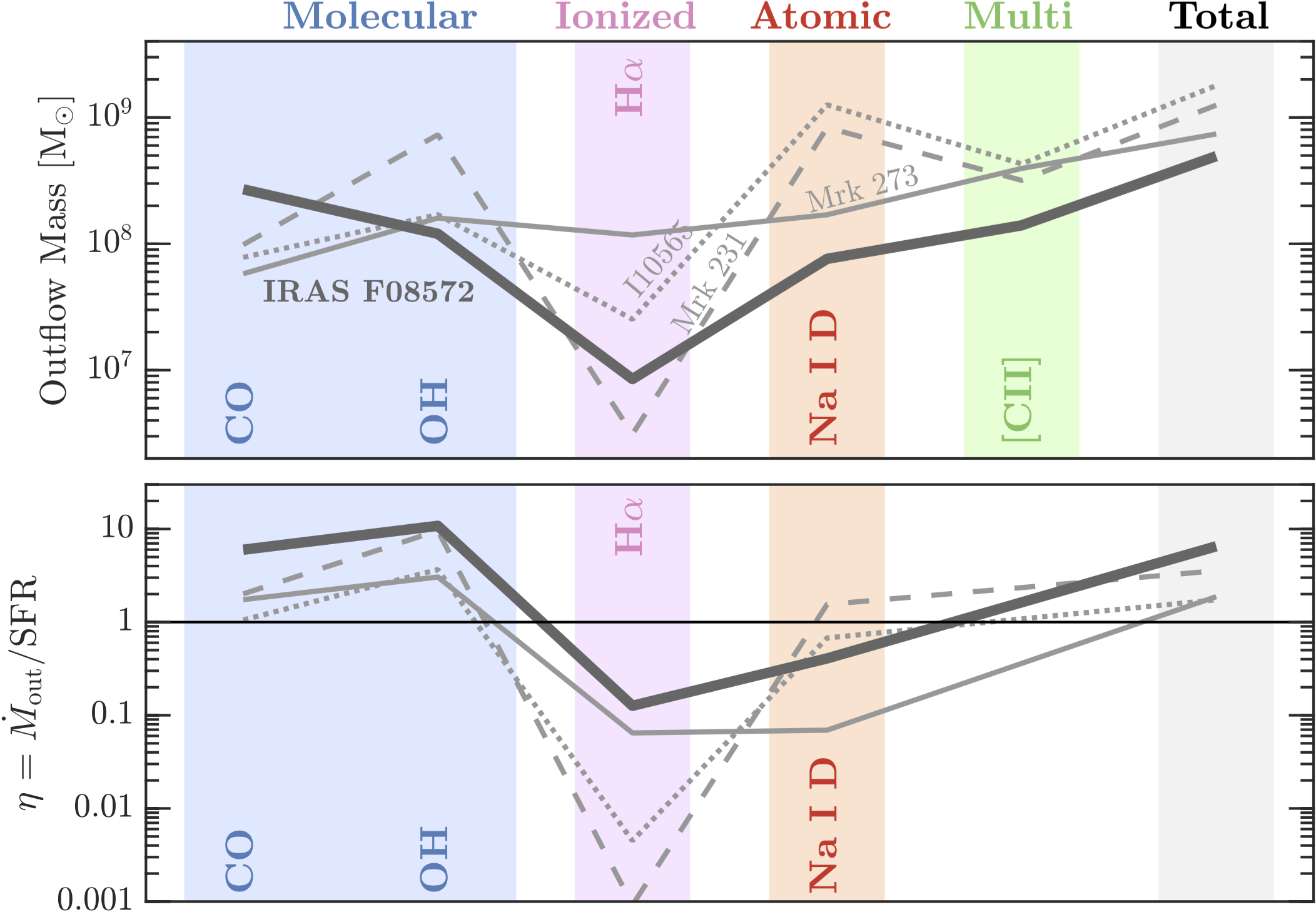}
\caption{Outflow mass (top) and mass loading factor $\eta$ (bottom) for different gas phases of the outflow, including molecular (as traced by CO and OH), ionized (\ha) and atomic phases (\nad), and most likely a combination of the three as traced by \cii. In the case of the sum of the phases (last column) we consider the molecular phase as traced by the CO line emission. We caution the reader that these measurements are affected by a series of assumptions on the physical conditions of the gas and geometry that can significantly impact the final value. The results for IRAS F08572+3915 are shown with a thick gray line. For comparison, outflow properties of other major merger, infrared luminous systems such as Mrk~231 (dashed line), Mrk~273 (solid line), and IRAS F10565+2448 are also shown \citep[][]{rhc_rupke13,rhc_veilleux13,rhc_cicone14,rhc_janssen16,rhc_gonzalez-alfonso17}.}
\label{fig:budget}
\end{figure*}

\section{Analysis}

\subsection{Comparison of the molecular, atomic, and ionized phases of the outflow}
\label{sec:other_phases}

The improved spatial resolution and sensitivity of the CO observations makes possible a comparison with other ISM phases of the outflow. Previously observed components of the wind include the atomic, ionized, and cold and warm molecular phases \citep[][]{rhc_sturm11,rhc_rupke13,rhc_rupke13b,rhc_janssen16,rhc_gonzalez-alfonso17}. Table \ref{tab:phases} summarizes the outflow properties of the different phases including the total mass in the outflow, the average (or typical) velocity, the outflow radius in kpc (if known), the mass outflow rate ($\dot{M}_{\text{out}}$), the momentum rate ($\dot{M}_{\text{out}}v$), and the energy rate ($1/2 \dot{M}_{\text{out}}v^{2}$). 

\medskip

\noindent {\it Cold molecular phase -- CO:} For a detailed description of the cold molecular phase of the outflow based on the CO(1-0) line see Section~\ref{sec:outflow_size}.

\medskip

\noindent {\it Molecular phase -- OH:} The molecular phase of the wind in the NW component of IRAS~F08572+3915 is also seen in multiple OH transitions \citep{rhc_sturm11,rhc_gonzalez-alfonso17}. It is important to note that the OH observations trace only the central part of the outflow: because the blueshifted wings in the OH lines are observed in absorption, a FIR continuum background is required in order to see the outflow. In IRAS~F08572+3915 this continuum has a half-light radius as small as 0.4~kpc at 70~$\mu$m, and 0.7~kpc at 100~$\mu$m \citep{rhc_lutz16}. This suggests that there is an overlap between the molecular outflow traced by the CO(1-0) and OH transitions. Note, however, that most likely the gas traced by the OH transitions is more sensitive to the nuclear outflowing gas \citep{rhc_gonzalez-alfonso17}, while CO(1-0) emission is more sensitive to the more extended, kiloparsec scale blue-shifted outflow. According to the model of the outflow by \citep{rhc_gonzalez-alfonso17}, the total molecular gas mass in the outflow as traced by the OH transitions is $M_{\rm mol,OH}=1.2\times10^8$~M$_{\odot}$, which is consistent, within the uncertainties, with the CO-based molecular gas mass of $M_{\rm mol,CO}=2.7\times10^8$~M$_{\odot}$. 

\medskip

\noindent {\it Warm molecular phase -- \hd:} \citet{rhc_rupke13b,rhc_rupke16} present OSIRIS/Keck  observations of the warm H$_{2}$ outflow, which has velocities between $-700$~km~s$^{-1}$ and $-1000$~km~s$^{-1}$. The outflow accelerates over a few hundred parsec, which suggests that the gradient in the CO outflows is undetected due to beam smearing. As Figure~\ref{fig:co_multi} shows, the CO(1-0) blueshifted wing is aligned with the warm H$_{2}$ outflow \citep{rhc_rupke16}. The H$_{2}$ outflow emerges along the minor axis of a very small-scale ($\lesssim500$~pc) disk that is oriented very differently from the large-scale disk as traced in the optical \citep{rhc_rupke13} or the cold molecular gas. The outflow mass and outflow rate in warm H$_{2}$ is $\sim 10^{-4}$ times the mass and outflow rate traced by CO(1-0). The same mass ratio between warm and cold H$_{2}$ was found in M82 by \citet{rhc_veilleux09c}. Note, however, that the H$_{2}$ observations only cover a small area of the whole outflow (the FOV is $1'' \times 2.9''$).

\medskip

\noindent {\it Neutral and ionized phases -- \nad\ and \ha:} The ionized phase of the outflow, as traced by broad blueshifted \ha\ line emission, reach very high velocities \citep[$\sim3300$~km~s$^{-1}$; ][]{rhc_rupke13} and extends along the major kinematic axis of the galaxy, similar to the blue wing of the molecular outflow. The atomic wind, traced by the \nad\ line in absorption, is offset from the nucleus by $\sim1-2$~kpc and partly overlaps with the blueshifted ionized and molecular winds. It reaches velocities up to $\sim1000$~km~s$^{-1}$ \citep[][]{rhc_rupke13}. The FoV of the observations of the atomic and ionized phases of the outflow covers red and blue components of the biconical outflow. Still, in both cases only the blueshifted part of the outflow is detected, most likely because the redshifted part is obscured by the disk.

\medskip

\noindent {\it Multi-phase -- \cii:} \cii~158~$\mu$m line emission can arise from the ionized, molecular, and atomic media \citep[e.g.,][]{rhc_pineda13,rhc_abdullah17,rhc_rhc17}. Therefore, the high-velocity ($\sim800$~km~s$^{-1}$) gas detected in the \cii\ spectrum of \name\ by \cite{rhc_janssen16} could be probing a combination of phases in the outflow. \cite{rhc_janssen16}, assuming that the gas in the outflow follows typical ULIRG-like conditions ($n=10^5$~cm$^{-3}$ and $T=100$~K), estimated a total mass in the outflow of $M_{\rm out}=1.4\times10^8$~M$_{\odot}$, which is comparable to that measured in the cold molecular phase using the CO and OH lines.

\medskip

Which gas phase dominates the mass and energetics of the outflow? In Figure~\ref{fig:budget} we compare the outflow mass and mass loading factor of \name\ as a function of gas phase, understanding that  this exercise is limited, among other factors, by the assumptions on the wind geometry and luminosity-to-mass conversion factors that can introduce up to an order-of-magnitude uncertainty. For comparison, we also include multi-phase outflow measurements for other three major merger, luminous infrared systems: Mrk~231, Mrk~273, and IRAS~F10565+2448 \citep{rhc_rupke13,rhc_veilleux13,rhc_janssen16,rhc_gonzalez-alfonso17}. We find that in \name\ the cold molecular gas is the dominant phase of the outflow. The atomic phase, however, is only a factor $\sim3$ lower than the molecular gas, which taking into account the uncertainties in the calculations, could be considered comparable. The same conclusion is true for the other three ULIRGs shown in the Figure.

The bottom panel of Figure~\ref{fig:budget} shows the mass loading factor (defined as $\eta=\dot{M}_{\rm out,mol}/{\rm SFR}$) measured in the different gas phases. We find that the mass loading factor in the molecular phase largely dominates over the atomic and ionized phase values, and it is the only phase where the rate of gas ejection is higher than the rate of molecular gas consumption, i.e., $\eta_{\rm mol}>1$. We discuss in more detail the implications of the high molecular mass outflow rate in Section~\ref{sec:depletion}.

\subsection{Star formation versus AGN activity as drivers of the main molecular outflow}\label{sec:power_source}

In this section we investigate if the nuclear starburst, the AGN, or a combination of the two are capable of driving the powerful molecular outflow observed in IRAS~F08572+3915. 

In an ideal scenario, outflows can be either momentum or energy driven \citep[e.g.,][]{rhc_f-g12,rhc_costa14}, which results in different predictions on how much the stellar and AGN feedback can contribute to the expansion of the wind. The observed high momentum boost in the molecular outflow of \name\ ($\sim20~L_{\rm AGN}/c$ or $\sim40~L_{\rm SF}/c$) can only be explained if the outflow is energy-driven. In that case,  energy injection by supernovae explosions and winds from massive stars is expected to be $\sim0.1-0.5\%$ of the starburst luminosity \citep[e.g.,][]{rhc_murray05,rhc_veilleux05}.  For IRAS~F08572+3915 this corresponds to $\sim(0.1-0.5)\%\times L_{\rm SB}\sim2-9\times10^{42}$~erg~s$^{-1}$, which is at least a factor of $\sim15$ lower than the measured molecular outflow kinetic luminosity (Eq.~\ref{eq:E}).If the main power source is the AGN, the maximum energy input is expected to be $\sim5\%$ of the AGN radiative power in case the coupling efficiency with the ISM is 100\% \citep[e.g.,][]{rhc_f-g12,rhc_zubovas12}. This results in $\sim5\%\times L_{\rm AGN}\sim2\times10^{44}$~erg~s$^{-1}$, which is comparable to the kinetic luminosity of the molecular outflow (Eq.~\ref{eq:E}). This suggests that the AGN is the main source driving the outflow, and that the coupling efficiency with the ISM is high as a result of a dense, thick and more spherical distribution of the gas and dust around the AGN. This scenario is consistent with the observed deeply dust obscured nature of the nuclear region in \name\ (See Section~1.1).

In summary, this simple and idealized analysis presented here suggests that the fast, kpc-scale molecular outflow in IRAS~F08572+3915 is energy conserving, driven by the AGN, and with a high ISM coupling efficiency.

\subsection{The origin of the fast gas blob $\sim6$~kpc away from the galaxy}\label{sec:origin_blob}

In addition to the main component of the outflow, we detect a gas blob of projected size $\sim1$~kpc and located $\sim6$~kpc north-west of the main system that is moving away at $\sim$900~km~s$^{-1}$ (see Section~\ref{sec:second_red_outflow} for details). Here we discuss two alternatives to explain its origin: a fossil outflow and a faint jet.

Outflow features resulting from episodic driving of the AGN are commonly known as fossil outflows, and are both expected from theory \citep[e.g.,][]{rhc_king11} and observed in nearby systems \citep[e.g.,][]{rhc_fluetsch19,rhc_lutz19}. In this scenario, the fast gas blob could be the result of an earlier phase of nuclear activity. For a distance of 6~kpc and assuming the gas has been driven out all the way from the center at a constant velocity of 900 km s$^{-1}$, the flow time from the center of the main galaxy is $\sim6$~Myr (this also assumes deprojection effects in velocity and radius cancel out). This rough estimate is consistent with the variability (or ``flickering") timescale of AGN activity of about $\sim0.1-1$~Myr \citep[e.g.,][]{rhc_schawinski15,rhc_king15b,rhc_zubovas16}, and the fact that outflow material can continue to expand for a time $\sim10$ times longer than the duration of the nuclear active phase \citep[][]{rhc_king11}.

The second alternative is that the gas blob is the product of the interaction between a relativistic jet and the ISM. Feedback by jets is mainly driven by ram and thermal pressure which results in outflows that are energy-conserving on all scales \citep[for a review see][]{rhc_wagner16}. As the jet opens its way through the clumpy galaxy disk and surrounding material it disperses atomic and molecular clouds in all directions. The interaction between the jet and the ISM can extend for kiloparsec scales \citep[e.g.,][]{rhc_morganti05,rhc_holt08,rhc_wagner11}.

There is no clear evidence for a jet  in \name. Old Very Large Array (VLA) radio observations by \cite{rhc_sopp91} tentatively detect a faint, extended structure in the north-south direction that extends for about 4\arcsec\ and could be interpreted as the signature of a faint jet. More recent observations with the upgraded Karl G. Jansky VLA by \cite{rhc_leroy11b} and \cite{rhc_barcos17} --with comparable angular resolution than \cite{rhc_sopp91}--, do not detect any extended component, only compact emission. Consistent with the scenario of no jet, the ratio between the rest-frame infrared and the 1.4~GHz monochromatic radio flux, $q_{\rm IR}$, indicates that \name\ is radio-quiet \citep[$q_{\rm IR}$ is 3.57, about an order of magnitude higher than the average value found in ULIRGs;][]{rhc_leroy11b}. We note, however, that there are examples of radio-quiet galaxies with jet-like winds indicating the existence of either a faint ongoing jet or a past jet event \citep[e.g.,][]{rhc_aalto16,rhc_fernandez-ontiveros19}.
Finally, the CO spectral line energy distribution (SLED) up to $J=11$ reveals highly excited gas in \name\ \citep[][]{rhc_papadopoulos10,rhc_pearson16}. While the strong AGN and starburst activity contribute significantly to the high molecular gas excitation, we cannot rule out that shocks resulting from a potential jet-dense ISM play a role in shaping the SLED beyond $J\approx7$ \citep[e.g.,][]{rhc_papadopoulos08,rhc_pellegrini13}. 

In summary, we do not have enough evidence to confirm there is or has been a faint radio jet operating in \name, but if it were, it opens the possibility for the fast gas blob to be the result of dense material accelerated by the jet far away from the nucleus. In that case the estimated flow time of $\sim6$~Myr in the fossil scenario would be obsolete.  A jet-driven outflow would be also consistent with the observed high momentum boost and energy conserving properties of the wind. Certainly, deeper and higher angular resolution observations are needed to confirm or rule out the jet-ISM interaction scenario.

\begin{figure}
\centering
\includegraphics[scale=0.13]{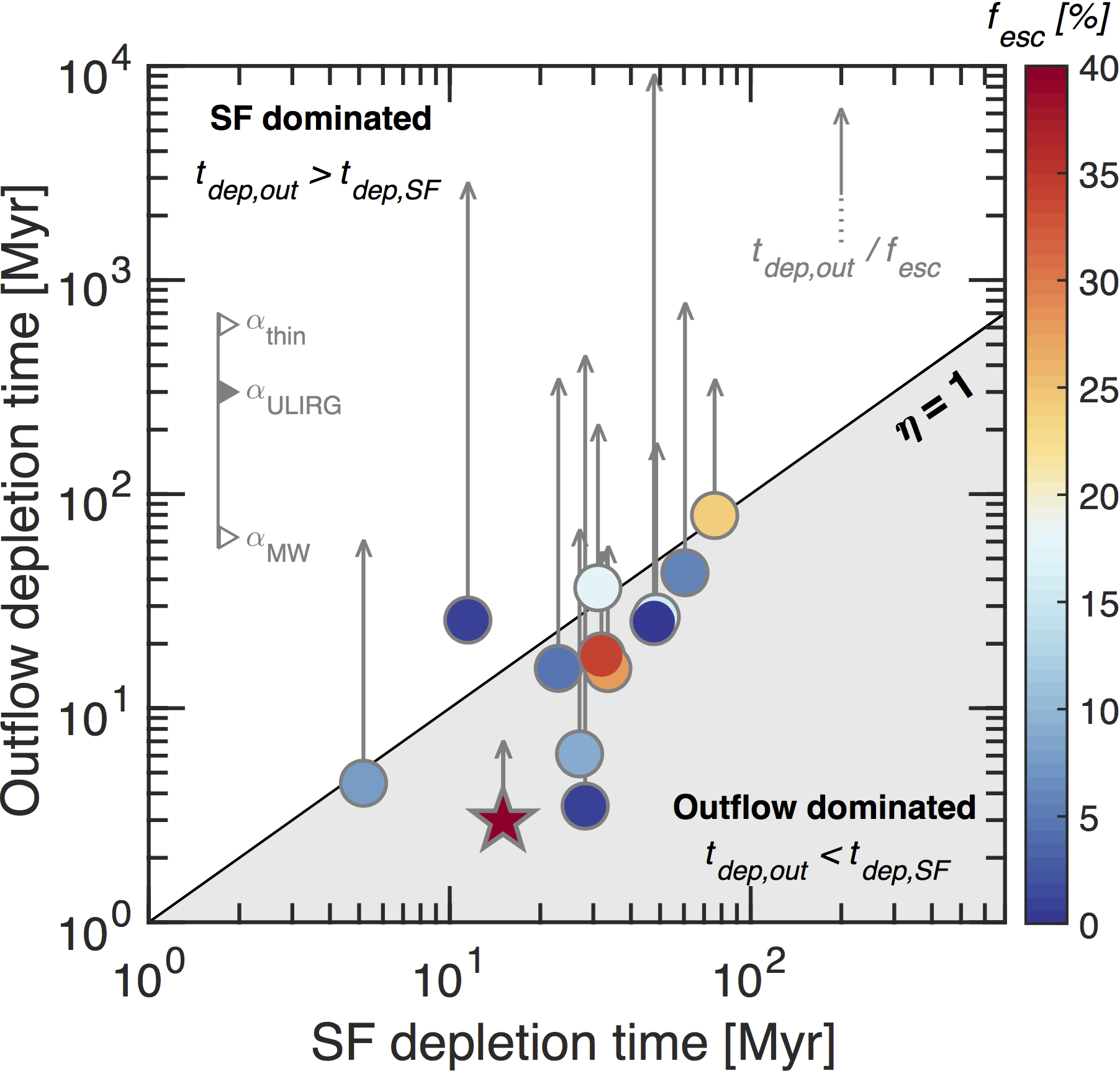}
\caption{Molecular gas depletion timescale (in Myr) due to gas consumption by star formation activity ($t_{\rm dep,SF}=M_{\rm mol}/{\rm SFR}$; abscissa) and gas removal by the outflow ($t_{\rm dep,out}=M_{\rm mol}/\dot{M}_{\rm out,mol}$; ordinate). IRAS~F08572+3915 is shown as a star, and other (U)LIRGs taken from the literature are shown as circles \citep{rhc_p-s18,rhc_fluetsch19}. The points are color-coded according to the escape fraction $f_{\rm esc}$, defined as the mass fraction of the molecular gas in the outflow that can escape the gravitational potential of its host. The diagonal line represents $t_{\rm dep,SF}=t_{\rm dep,out}$, or equivalently, a mass loading factor of $\eta=1$. The vertical lines show where the data points would move if we calculate the depletion time based on the rate of molecular gas ejected by the outflow that can escape the gravitational potential of its host (i.e., $\dot{M}_{\rm esc,mol}=f_{\rm esc}\times\dot{M}_{\rm out,mol}$.)} 
\label{fig:mass_loading}
\end{figure}

\subsection{AGN feedback and quenching of star formation}
\label{sec:depletion}

Quantifying the impact of the outflow on the star formation activity of IRAS~F08572+3915 is a very complicated problem that requires detailed knowledge on the accretion of fresh and/or recycled gas, the ejection of molecular gas by the outflow, how much of that gas can permanently escape from the galaxy, and the duty cycle of the AGN. Unfortunately, some of these key pieces are missing, and in this section we can only hypothesize on the fate of IRAS~F08572+3915 based on the measurements we have of the molecular gas reservoir, the star formation activity, and the energetics of the outflow. 

We start by comparing the time it would take the star formation and the outflow to exhaust --via consumption or ejection-- the total reservoir of molecular gas. The star formation depletion timescale, defined as $t_{\rm dep,SF}=M_{\rm mol}/{\rm SFR}$, is $t_{\rm dep,SF}\approx15$~Myr, which is at the low end of the range of $t_{\rm dep,SF}$ measured in (U)LIRGs  \citep[e.g.,][]{rhc_cicone14,rhc_gonzalez-alfonso17,rhc_shangguan19}. The molecular mass loading factor in IRAS~F08572+3915 is $\eta_{\rm mol}=\dot{M}_{\rm out,mol}/{\rm SFR}\approx5$, so the depletion time due to the outflow, $t_{\rm dep,out}=M_{\rm mol}/\dot{M}_{\rm out,mol}$, is only $\approx3$~Myr. This timescale reduces in half if we only consider the molecular gas in the nuclear $\sim1.5$~kiloparsec region (i.e., $t^{1.5~\rm kpc}_{\rm dep,out}\sim1.5$~Myr). Based on this short depletion timescale one could expect that the outflow is rapidly quenching the star formation activity in (at least) the central region of IRAS~F08572+3915. Before jumping to such conclusion, however, it is important to keep in mind that: (1) the AGN is variable and ``flickers'' on expected timescales of $\sim0.1-1$~Myr \citep[e.g.,][]{rhc_schawinski15,rhc_zubovas16}, and (2) an important fraction of the molecular gas that is ejected via the outflow could be later re-accreted and become available to fuel future episodes of star formation \citep[see for example the molecular outflows studied by][]{rhc_p-s18,rhc_fluetsch19,rhc_rhc19}. 

To obtain a rough estimate of the amount of molecular gas in the outflow of IRAS~F08572+3915 that can escape the gravitational potential of its host we need to determine the escape velocity from the system. We start by calculating the dynamical mass of the main (or northern) component of \name. For this we use {\it dysmalpy}, which is an updated version of the dynamical model code {\it dysmal} \citep{rhc_cresci09,rhc_davies11} that now includes an MCMC sampling procedure. We model the CO velocity field using as a free parameters the dynamical mass ($M_{\rm dyn}$), effective radius of an exponential disk ($R_{\rm eff}$), and inclination ($i$) of the galaxy. The details of the kinematic analysis are discussed in Appendix~A. As we show in Figure~\ref{fig:residual}, the {\it dysmal} model of the velocity field does a good job reproducing the bulk rotation of the system. From the MCMC sampling of the joint posterior probability distributions of the model parameters (see Figure~\ref{fig:corner_plot}) we determine that the dynamical mass and the effective radius are $log_{10}(M_{\rm dyn}/{\rm M_{\odot}})=10.19^{+0.13}_{-0.34}$ and $R_{\rm eff}=1.04^{+0.17}_{-0.23}$~kpc, respectively. Following a similar approach to \cite{rhc_fluetsch19}, assuming a Hernquist profile for the density \citep{rhc_hernquist90} we estimate an escape velocity from the gravitational potential of the galaxy at $\sim R_{\rm eff}$ of $v_{\rm esc}\approx850$~km~s$^{-1}$. If we then integrate the CO spectrum of the main galaxy in the range where velocities are higher than the escape velocity, we estimate that the global fraction of molecular gas that can escape the system is $f_{\rm esc}\approx0.4$. This value is at the high end of the distribution of global escape fractions of molecular gas computed for other ULIRGs by \cite{rhc_p-s18} and \cite{rhc_fluetsch19}. It is also consistent with the high molecular escape fraction of $f_{\rm esc}\approx0.25$ derived based on the analysis of the OH transitions by \cite{rhc_gonzalez-alfonso17}. One caveat worth mentioning in this simplified calculation is that due to the limited spatial resolution and the lack of precise knowledge concerning the wind geometry, it is impossible to estimate what fraction of the outflowing gas will escape at distances $R\lesssim R_{\rm eff}$. Taking this into account would most likely reduce the global escape fraction.

If we fold in the outflow escape fraction into the calculation of the mass loss rate we obtain the molecular gas mass {\it escape} rate, which for \name\ is $\dot{M}_{\rm esc,mol}=f_{\rm esc}\times\dot{M}_{\rm out,mol}\approx150$~$M_{\odot}$~yr$^{-1}$. This implies that the time it would take for the outflow to remove the molecular gas from the galaxy gravitational potential is only $\sim3$~Myr for the inner $\sim1.5$~kiloparsec region, and $\sim7$~Myr for the whole molecular content of the system. Figure~\ref{fig:mass_loading} put these timescales in context with those measured in other (U)LIRGs by \cite{rhc_fluetsch19} and \cite{rhc_p-s18}. The color symbols represent the star formation and outflow depletion timescales if we consider all the gas that is being ejected.

All of the (U)LIRGs except two (IRAS 14348 NE and PG 0157+001) have $t_{\rm dep,out}\lesssim t_{\rm dep,SF}$ (or equivalently $\eta\gtrsim1$), i.e., the depletion of the molecular gas in these systems is dominated by the outflow. 
This scenario drastically changes if we now only consider the gas in the outflow that is fast enough to escape the gravitational potential of the galaxy. This change increases the outflow depletion timescales by a factor $f_{\rm esc}^{-1}$, and the new position of the galaxies is shown with vertical grey lines. We note that all the galaxies, except IRAS~F08572+3915, have shorter depletion timescales associated to the starburst activity, not the outflow ($t_{\rm dep,SF}\lesssim t_{\rm dep,out}$, or equivalently $\eta_{\rm esc}=\eta\times f_{\rm esc}\lesssim1$). 
In the case of IRAS~F08572+3915, the rate of gas ejection that can escape is a factor of two higher than the rate of gas consumption. This result is consistent with the low molecular gas content measured in this galaxy relative to other (U)LIRGs: it is the system with the lowest molecular gas content in the sample of \cite{rhc_solomon97}, and the one with the highest $L_{\rm FIR}/M_{\rm H_2}$ ratio in the sample of \cite{rhc_gonzalez-alfonso15}.

The fact that the mass loading factor in \name\ is higher than one --even after we consider only the gas that is fast enough to escape the system-- shows the potential the outflow has to deplete the molecular gas from the central region and thus prevent future episodes of star formation \cite[e.g.,][]{rhc_hopkins10,rhc_zubovas12,rhc_zubovas17}. It is important to keep in mind, however, a series of factors that complicate this first order interpretation. For one we have AGN variability, that will make the actual ejection time of molecular gas longer. Another important unknown is the fraction of gas that is removed from the gravitational potential of the host that could be reaccreted at later times.

\section{Summary and conclusions}
\label{sec:conclusions}

In this paper we present deep and spatially resolved observations of the molecular gas in the ultra-luminous infrared galaxy \name\ based on new, deep ($\sim$50~hours) CO line observations with the NOEMA interferometer. This system is known to host a powerful, multi-phase AGN-driven outflow \citep{rhc_sturm11,rhc_rupke13,rhc_rupke13b,rhc_cicone14,rhc_janssen16,rhc_gonzalez-alfonso17}. The goal of this work was to characterize in detail the molecular phase of the wind and explore its impact on the star formation activity. 

We highlight the following points:

\begin{enumerate}
\item Compared to previous observations of the CO(1-0) line emission by \cite{rhc_cicone14}, our data achieves a better spatial resolution ($\theta=1.4\arcsec\times1.3\arcsec$ versus $\theta=3.1\arcsec\times2.7\arcsec$) and sensitivity ($\sigma=0.06$~mJy~beam$^{-1}$ versus $\sigma=0.2$~mJy~beam$^{-1}$). This allow us to spatially-resolve the molecular outflow in the main galaxy, and to detect for the first time the molecular gas in the minor galaxy of the interacting pair.
\item The molecular outflow in \name\ is fast ($v_{\rm out}\approx1200$~km~s$^{-1}$), massive ($M_{\rm mol,out}\approx1/4\times M_{\rm mol,disk}$), and most likely has a biconical shape with a wide opening angle. No velocity gradient in the outflow is observed. 
\item We detect an additional outflow component in the receding side that is detached from the biconical structure. This ``gas blob'' has a molecular gas mass of $\approx5\times10^7$~M$_{\odot}$, is located at about 6~kpc from the main galaxy, and is moving away at $\sim900$~km~s$^{-1}$. Its origin could be associated to the intermittent (or ``flickering'') nature of AGN activity or the potential existence of a faint jet that interacted with the surrounding dense ISM medium.
\item Compared to other gas phases in the outflow (warm molecular, ionized, and atomic), the cold molecular phase dominates both the outflow mass and the mass loss rate in \name. In fact, this is the only phase where the mass loading factor $\eta$ is greater than unity ($\eta_{\rm cold,mol}>1>\eta_{\rm neutral}>\eta_{\rm ion}$).
\item The fraction of molecular gas in the outflow that can escape the gravitational potential of the galaxy is $f_{\rm esc}\sim0.4$, which is at the high end of the range of escape fractions measured in other (U)LIRGs \citep[$f_{\rm esc}\sim0.01-0.3$;][]{rhc_p-s18,rhc_fluetsch19}.
\item The mass outflow rate of high-velocity gas that can {\it escape} the galaxy (i.e., $\dot{M}_{\rm esc,mol}=f_{\rm esc}\times\dot{M}_{\rm out,mol}$) is $\approx150$~$M_{\odot}$~yr$^{-1}$, which is a factor of $\sim2$ higher than the SFR. Compared to the samples of (U)LIRGs in \cite{rhc_p-s18} and \cite{rhc_fluetsch19}, \name\ is the only system with a powerful enough outflow to deplete the central molecular gas by ``blowing it away'' on a timescale shorter than that of star formation.
\end{enumerate}

\begin{acknowledgements}
We thank the referee, K. Dasyra, for very useful comments and suggestions which greatly improved the manuscript.  R.H-C. would like to dedicate this work in memory of his father-in-law, Fernando Antonio Bravo Parraguez (1966-2019). A.J. and R.H-C. would like to thank Roberto Neri, Michael Bremer and the rest of the IRAM staff for their help with the data calibration, and Loreto Barcos for helpful discussions about VLA observations of the source. R.H.C. Figure~\ref{fig:hst_co} is based on observations made with the NASA/ESA Hubble Space Telescope, and obtained from the Hubble Legacy Archive, which is a collaboration between the Space Telescope Science Institute (STScI/NASA), the Space Telescope European Coordinating Facility (ST-ECF/ESA) and the Canadian Astronomy Data Centre (CADC/NRC/CSA). S.V. acknowledges support from a Raymond and Beverley Sackler Distinguished Visitor Fellowship and thanks the host institute, the Institute of Astronomy, where this work was concluded. S.V. also acknowledges support by the Science and Technology Facilities Council (STFC) and by the Kavli Institute for Cosmology, Cambridge. RM acknowledges ERC Advanced Grant 695671 ``QUENCH'' and support by the Science and Technology Facilities Council (STFC).  E.GA is a Research Associate at the Harvard-Smithsonian Center for Astrophysics, and thanks the Spanish Ministerio de Econom\'{\i}a y Competitividad for support under project ESP2017-86582-C4-1-R. CC acknowledges funding from the European Union's Horizon 2020 research and innovation programme under the Marie Sklodowska-Curie grant agreement no. 664931.

\end{acknowledgements}


\begin{appendix}

\section{CO(1-0) spectrum of the fast gas blob from the A+B array configuration observations}\label{blob_spec_AB}

Figure~\ref{fig:gasblob_spec} shows the CO(1-0) spectrum extracted within a $R=1.5\arcsec$ circular aperture centered on the the second redshifted outflow component ($\alpha=09:00:25.2$, $\delta=+39:03:58.8$) located $\sim$6~kpc north of the NW galaxy. The diference between the red and grey spectra is that the former is based on A+B array configuration data, while he latter is extracted in the combined A+B+C+D data, identical to that shown in Figure~\ref{fig:gasblob_spec}. We confirm that the fast gas blob is detected in the A+B only data.

\begin{figure}
\centering
\includegraphics[scale=0.14]{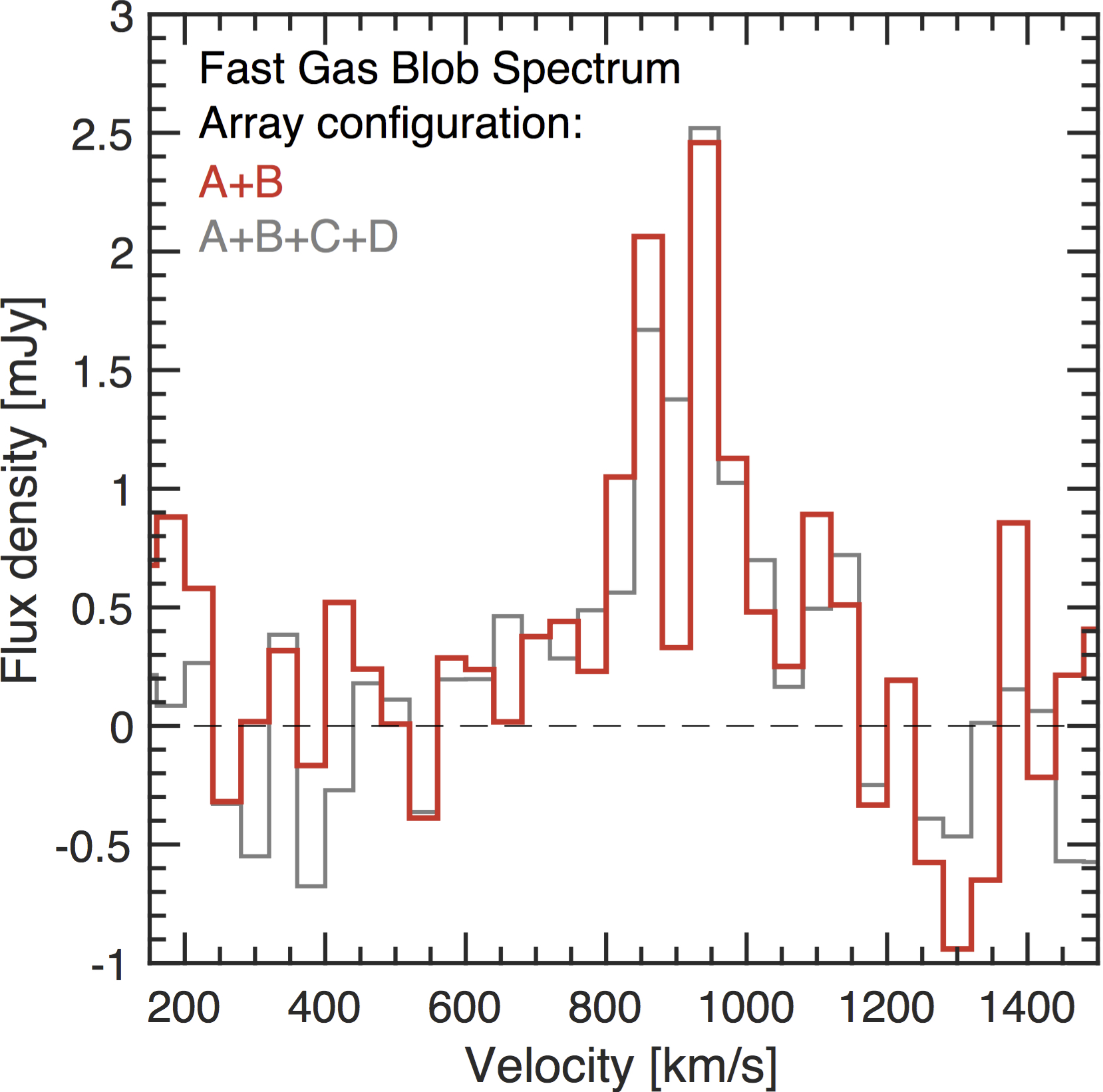}
\caption{CO(1-0) spectrum of the outflowing gas blob located $\sim$6~kpc north of the NW galaxy and that is independent of the main outflow structure. In red we show the spectrum extracted from the A+B only data, while in grey we show the spectrum from the combined A+B+C+D data (which is identical to the one shown in Figure~\ref{fig:gasblob_spec}).
}
\label{fig:gasblob_spec_AB}
\end{figure}

\section{{\it dysmalpy} modeling of the kinematics}

We model the 2D velocity field of the body ($\pm400$~km~s$^{-1}$) of \name\ to constrain its dynamical mass, effective radius, and inclination using an updated version of the dynamical fitting code {\it dysmal} \citep{rhc_cresci09,rhc_davies11,rhc_wuyts16,rhc_uebler18}. The code creates a three-dimensional mass model of the galaxy which is then compared to the data based on an  implementation of an MCMC sampling procedure using the EMCEE package \citep{rhc_foreman-mackey13}. One of the advantages of using {\it dysmalpy} is that it accounts for beam-smearing effects by convolving with the two-dimensional PSF (or beam) of the galaxy.

Free parameters in our modeling are the dynamical mass ($M_{\rm dyn}$), the effective radius of an exponential disk ($R_{\rm eff}$), and the inclination ($i$). For these parameters we choose Gaussian priors which reflect our prior knowledge about their values and uncertainties. For the dynamical mass we chose the range $M_{\rm dyn}=[10^{9},10^{11}]~M_{\odot}$ \citep[a previous estimate of the dynamical mass derived from H$\alpha$ line kinematics yielded $M_{\rm dyn}\approx10^{10}~M_{\odot}$;][]{rhc_arribas14}, for the effective radius we chose the range $R_{\rm eff}=[0.5,1.5]$~kpc, based on the range of values measured from HST F814W and F160W data \citep{rhc_garcia-marin09}, and it is also consistent with our CO measurement (see Table~\ref{tab:components}). From visual inspection of the HST data we set the boundaries for the disk inclination $i$ prior between 20 and 60 degrees. Fixed parameters in our modeling are the position angle and the central position of the velocity field which we set to 120$^{\circ}$ and $\alpha=09:00:25.3, \delta=+39:03:54.2$, respectively, based on visual inspection.

\begin{figure*}
\centering
\includegraphics[scale=0.038]{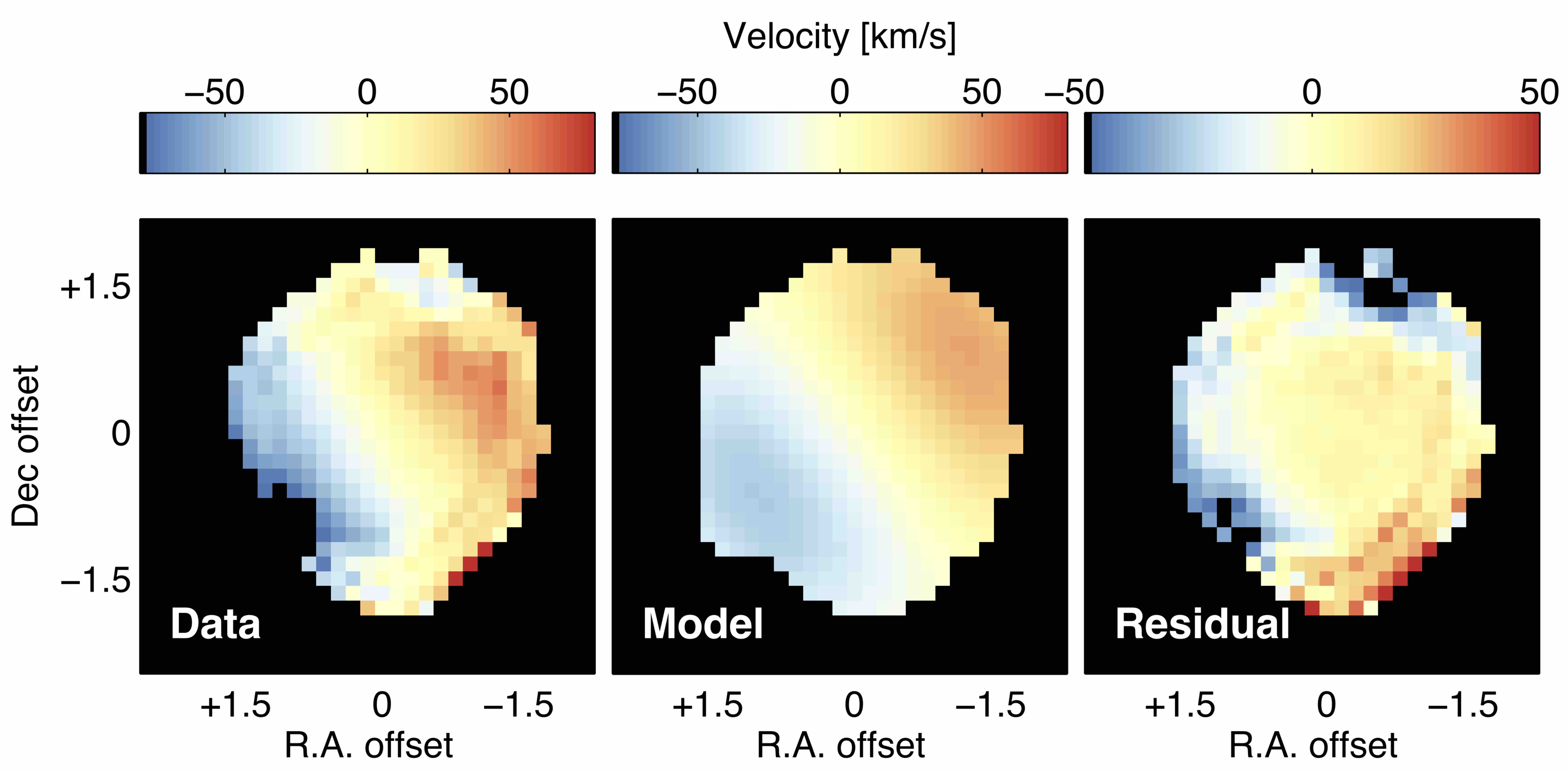}
\caption{From left to right, CO(1-0) velocity field of \name, the best fit model from {\it dysmalpy}, and the residual map between the two. The observed motions in \name\ are not purely rotational, as expected from a system that is undergoing a major interaction and host a powerful nuclear outflow. Nonetheless, the best fit velocity field from the {\it dysmalpy} code does a reasonable job reproducing the major kinematics features of \name.}
\label{fig:residual}
\end{figure*}

Figure~\ref{fig:residual} shows the CO(1-0) velocity field of \name, the velocity field extracted from the ${\it dysmalpy}$ model cube, and the residual map. Overall, the kinematic model does a good job reproducing the bulk of the rotational motion observed in CO ---the amplitude of the residual throughout a large portion of the disk is $\lesssim15$~km~s$^{-1}$. Near the edges, however, the model fails to capture the S-shaped pattern in the kinematics of \name, which is a signature of non-circular orbits and indicate deviations of the gravitational potential from axisymmetry or possible outflows and inflows \citep[e.g.,][]{rhc_roberts79,rhc_wong04}. This is expected as the ${\it dysmalpy}$ model does not include any gas inflow or outflow component.

\begin{figure*}
\centering
\includegraphics[scale=0.2]{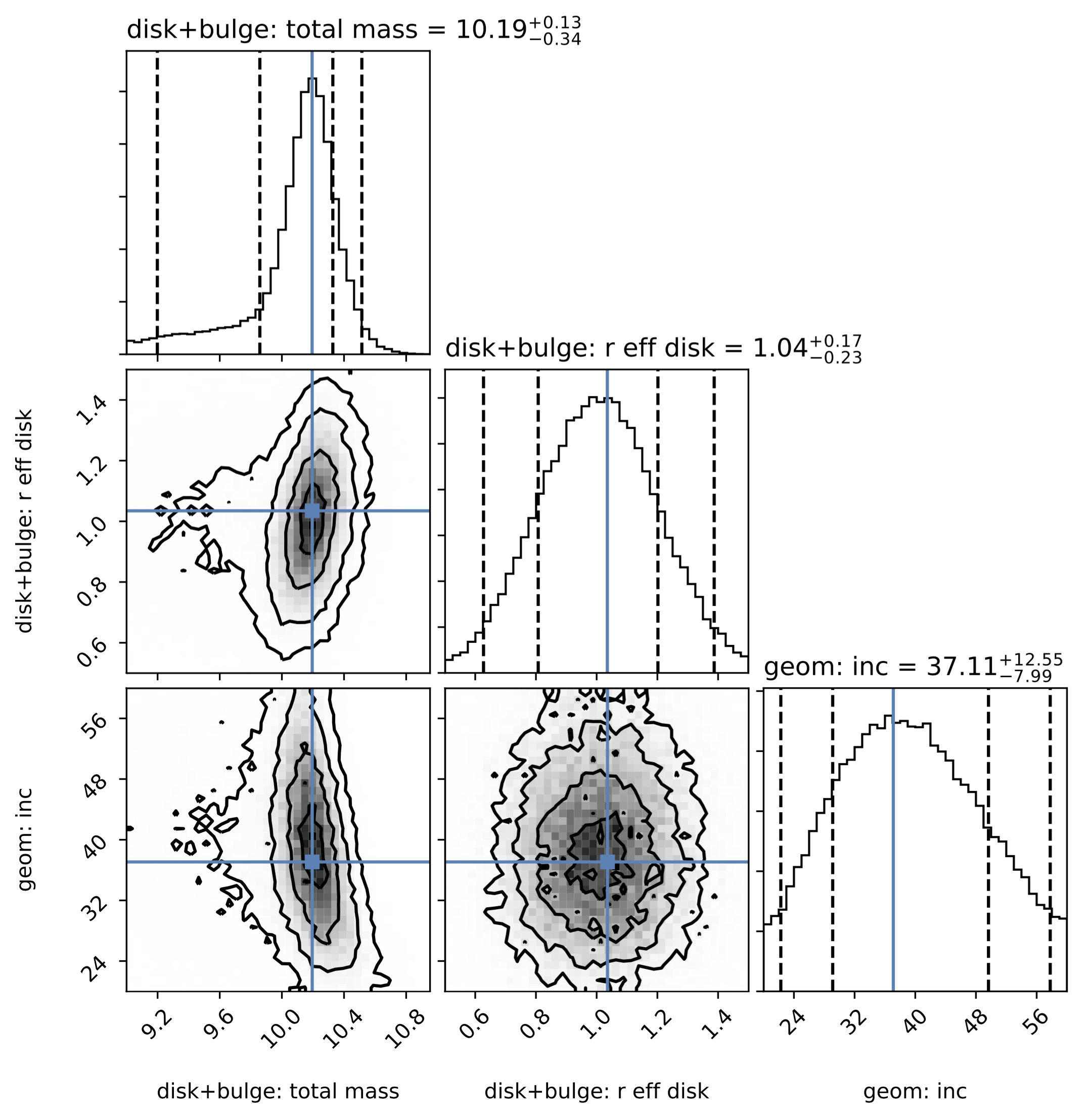}
\caption{MCMC ``corner plot'' of our best-fit parametric model (see Figure~\ref{fig:residual}). The figure shows the one- and two-dimensional projections of the posterior probability distributions of the three free parameters: the dynamical mass ($M_{\rm dyn}$), the effective radius ($R_{\rm eff}$) and the inclination ($i$). The median values and 1$\sigma$ confidence ranges of the marginalized distributions are indicated by the dashed lines in the 1D histograms. The median values are also shown as blue squares on top of the 2D histograms. The contours show the 1, 2, and 3$\sigma$ confidence levels of the 2D distributions.}
\label{fig:corner_plot}
\end{figure*}

Figure~\ref{fig:corner_plot} shows the MCMC sampling of the joint posterior probability distributions of the model parameters (or the MCMC ``corner plot''). Because the posterior distribution is well behaved, we choose our fiducial model to be represented by the median values of the individual marginalized distributions (blue lines), with uncertainties represented by the 1$\sigma$ confidence ranges (dashed lines). Thus, our analysis based on the ${\it dysmalpy}$ model suggests that $log_{10}(M_{\rm dyn}/{\rm M_{\odot}})=10.19^{+0.13}_{-0.34}$, $R_{\rm eff}=1.04^{+0.17}_{-0.23}$~kpc, and $i=37.11^{+12.55}_{-7.99}$

\end{appendix}


\bibliographystyle{aa}
\bibliography{references.bib}

\end{document}